\documentclass[aps,prl,twocolumn,superscriptaddress,amsmath,amssymb]{revtex4-2}

\usepackage[T1]{fontenc}
\usepackage{graphicx}      
\usepackage{dcolumn}       
\usepackage{bm}            
\usepackage{mathtools}     
\usepackage{amsfonts}
\usepackage{upgreek}
\usepackage{gensymb}
\usepackage{stmaryrd}
\usepackage{mathrsfs}
\usepackage{dsfont}
\usepackage{accents}
\usepackage[version=4]{mhchem}
\usepackage{xcolor}
\usepackage{booktabs}
\usepackage{array}
\usepackage{multirow}
\usepackage{makecell}
\usepackage{placeins}
\usepackage[inline]{enumitem}

\newcommand{\desc}[1]{\parbox[t]{12cm}{\raggedright\strut#1\strut}}

\definecolor{xlinkcolor}{cmyk}{1,1,0,0} 
\usepackage[
    colorlinks=true,
    linkcolor=xlinkcolor,
    citecolor=xlinkcolor,
    filecolor=black,
    urlcolor=xlinkcolor
]{hyperref}

\begin{document}

\preprint{APS/123-QED}

\title{\textbf{Quantum Noise-Limited Terahertz Interferometry for Distinguishing\\ Ultra-Compact Supermassive Galactic Objects} 
}%

\author{Vishwas Patel}
\email{vrhpatel2000@gmail.com (VP)}
\affiliation{Space Applications Center, Indian Space Research Organisation, Ahmedabad, India-380015}
\author{P. S. Joshi}
\email{pankaj.joshi@ahduni.edu.in (PSJ)}
\affiliation{International Centre for Space and Cosmology, School of Arts and Sciences, Ahmedabad University, Ahmedabad-380009, India}%

\author{Swalih A. Valiyakath}
\email{swalihahamed10@gmail.com (SAV)}
\affiliation{International Centre for Space and Cosmology, School of Arts and Sciences, Ahmedabad University, Ahmedabad-380009, India}
\altaffiliation{Indian Institute of Science Education and Research Mohali, Sector 81 SAS Nagar, Manauli PO 140306, Punjab, India}

\author{Munn V. Shukla}
\affiliation{Space Applications Center, Indian Space Research Organisation, Ahmedabad, India-380015}

\author{Mehul R. Pandya}
\affiliation{Space Applications Center, Indian Space Research Organisation, Ahmedabad, India-380015}



\date{\today}

\begin{abstract}
The images of the galactic centers, M87$^*$ and Sgr~A$^*$, obtained by the Event Horizon Telescope (EHT) provide hitherto unknown revelations on these supermassive objects. The key issue here is the presence or otherwise of an event horizon at these ultra-compact entities.  We show here that distinguishing the central images generated by gravitational lensing of a Schwarzschild black hole and a Joshi--Malafarina--Narayan (JMN-1) central naked singularity provides important insights into the nature of these supermassive objects and would help test the cosmic censorship conjecture, central to black hole physics. For optically thin spherical accretion onto Sgr~A$^*$, their interferometric visibilities become separable, independently of the emission model, on three baseline ranges within
1.4--6.2\,G$\lambda$ at errors of $\simeq36\,\mu$Jy, and at $L_\star=2.64\,$G$\lambda$ at $\simeq71\,\mu$Jy. This is about 28 times below the 2017 Event Horizon Telescope statistical uncertainty. We suggest that the widest detection window demands minutes-long coherent integrations, favoring a near-quantum-limited space interferometer.
\end{abstract}



\maketitle

\section{Introduction}

Horizon-scale imaging of supermassive galactic central objects has turned the possible existence of an event horizon there into an observational question. The Event Horizon Telescope (EHT) has resolved a ring of emission encircling a central brightness depression in M87$^*$~\cite{EHT2019a}, and also for the  Sgr~A$^*$~\citep{EHT2022_SgrA_II,EHT2022_SgrA_III}, which is consistent with a predicted black-hole shadow ~\citep{akiyama2022first,event2022first}. The shadow, however, is set by the photon sphere, and in general by the nature of the central singularity  \citep{joshi2020shadow}, rather than due to the event horizon itself. Therefore, horizonless compact objects with photon spheres, such as naked singularities formed in gravitational collapse~\cite{joshi2011equilibrium}, wormholes~\cite{bambi2013can,jusufi2022}, and gravastars~\cite{gravastar}, could also cast shadows of the same size as a black hole would create ~\cite{bambi2013can,shaikh2019shadows,vagnozzi2023}. Among non-spinning alternatives, the Joshi--Malafarina--Narayan type-1 (JMN-1) naked singularity~\cite{joshi2011equilibrium} is one of the most compelling black-hole mimickers, as its shadow size remains consistent with the Sgr~A$^*$ observations~\cite{akiyama2022first}.

At stake here is the Cosmic Censorship Conjecture (CCC), which states that gravitational collapse of massive matter clouds always cloak their final state singularity behind an event horizon of gravity~\cite{penrose1969}. This  remains unproved for more than last five decades, and is directly challenged by several naked-singularity collapse models~\cite{joshi2011equilibrium}. It would seem, however, that ongoing recent observational developments  could make a test of CCC possible. The next-generation EHT has set its science requirements~\cite{johnson2023ngeht}, space-based submillimeter VLBI missions such as the Black Hole Explorer are under development~\cite{bhex2024}, and phased arrays~\cite{almaArea} along with efforts towards near-quantum-limited terahertz receivers~\cite{endfireSIS,malnou2021prxq,kitwta} continue to lower VLBI noise. Yet, when JMN-1 hosts a photon sphere, no definite measurable signatures are known that discriminate it at a finer level from a Schwarzschild black hole of equal mass, either theoretically or observationally. In the absence of such signatures, the black-hole nature of Sgr~A$^*$ remains an assumption rather than a measurement, and forthcoming arrays cannot be designed to meet the requirements of a censorship test.

In fact, earlier comparisons appear to suggest that the images of the two spacetimes are nearly indistinguishable whenever JMN-1 hosts a photon sphere ($2/3\le M_0<4/5$). It was seen that thin Novikov--Thorne disks truncate outside the photon sphere and mimic black-hole images~\cite{shaikh2019can}; semi-analytic radiatively inefficient flows render JMN-1 slightly brighter than the
Schwarzschild BH, thus showing a small but non-vanishing difference in their intensity distributions~\cite{saurabh2024}. The spherical accretion setups, similar to those adopted in this work, yield shadows of identical radius, with the difference increasing up to the shadow boundary~\cite{kaur2021comparing, shaikh2019shadows}.
In most other scenarios, larger luminosity and spectral differences arise for collapse-formed naked singularities~\cite{joshi2014distinguishing}, whereas interferometric signatures have been previously computed for spacetimes with multiple photon spheres that differ from those considered here~\cite{chen2024interferometric}. The distinction between their shadows has hitherto been described only in the image plane, without any systematic analysis of their relative contrast variation across different emission models, inner emission cut-off boundaries, or metric parameters. Additionally, no study has analyzed their synthetic very long baseline interferometric (VLBI) signatures, which have been proposed as the next step~\cite{saurabh2024}. Their visibility magnitude differences, important baselines that carry them, and the required instrument sensitivity therefore remain unknown.

In this Letter, our purpose therefore is to quantify the Schwarzschild--JMN-1 differences in the intensity
and visibility domains and translate the same  into instrument requirements for
Sgr~A$^*$. We ray-trace both these spacetimes~\cite{schwarzschild1916,joshi2011equilibrium} under
optically thin Bondi--Michel accretion~\cite{bondi1952spherically,michel1972accretion}
for three emissivity laws and, in the Schwarzschild case, six inner emission radii
from the horizon to the ISCO~\cite{kocherlakota2022}. We then compare the images
through their relative contrast and several image-quality metrics. We then also
compute the visibilities, define a guaranteed difference that holds for every image
extent consistent with the observed size of Sgr~A$^*$, and derive the instrument
error and baselines required to detect this difference against the 2017 EHT
statistical uncertainties and the refractive floor.

Spherical accretion makes the analysis independent of the viewing inclination and
reduces the two-dimensional visibility transform to a one-dimensional Hankel
transform, while placing the shadow edge at the photon
ring~\cite{narayan2019shadow}. Since the geometries coincide outside $R_{\rm b}$,
the difference arises from rays with $\xi<3\sqrt{3}M$ that probe the JMN-1 interior.
In the Schwarzschild case, the inner emission cut-off radius $R_{\rm in}$ is a free parameter of the emission model~\cite{kocherlakota2022}, whereas the JMN-1 emission extends inward to the singularity till fixed $R_{\rm in} = 10^{-3}M$. Hence, for $R_{\rm in}>3M$, the emission of JMN-1 between the photon sphere and $R_{\rm in}$ also contributes, extending the difference beyond the shadow edge. Working with visibilities avoids image-reconstruction priors. For all models, we quote only the smallest difference over all image truncations which are consistent with the observed source size, so the requirements hold irrespective of truncation choice.

Independently of the emission model, the two spacetimes become separable on
$1.43$--$2.34$, $2.55$--$3.12$, and $5.94$--$6.23\,\mathrm{G}\lambda$ once the
per-visibility error reaches $\simeq36\,\mu$Jy, and on the single most favourable
baseline, $L_\star=2.64\,\mathrm{G}\lambda$, at $\simeq71\,\mu$Jy, about 28 times
below the 2017 EHT statistical uncertainty. When the emission begins far from the
horizon, the difference is already detectable at the 2017 sensitivity and exceeds
the refractive-scattering floor in a single epoch, whereas emission reaching closer
to the horizon additionally demands averaging over many epochs. Opening the widest
detection window requires coherent integrations of several minutes, attainable in
principle only with a near-quantum-limited space interferometer.

The Appendices derives the inner emission radii from the EHT ring sizes (Appendix~A), details the image-extent calibration (Appendix~B), and derives the station sensitivity requirements (Appendix~C). The Supplemental Material contains additional contrast maps, the dependence on $M_0$, the sensitivity requirements and prominent baseline windows for all analyzed models, the image-metric analysis, and the proposed space-based architecture. Our Python-based ray-tracing and visibility codes are publicly available~\cite{patel_2026_zenodo}.

\section{Spacetimes and relativistic shadows}\label{sec:spacetimes}
We compare the Schwarzschild black hole (BH) \citep[][]{schwarzschild1916} and the Joshi--Malafarina--Narayan
type-I (JMN-1) naked singularity \citep{joshi2011equilibrium}. In geometric
units ($G=c=1$), the line element for both can be written as
\begin{equation}
  \mathrm{d}s_i^2 = -f_i(r)\,\mathrm{d}t^2 + \frac{\mathrm{d}r^2}{g_i(r)}
  + r^2\,\mathrm{d}\Omega^2 ,
\end{equation}
where, for Schwarzschild ($i=0$),
\begin{equation}
  f_0 = g_0 = 1-\frac{2M}{r},
\end{equation}
and for JMN-1 ($i=1$),
\begin{align}
  f_1 &= (1-M_0)\left(\frac{r}{R_{\rm b}}\right)^{M_0/(1-M_0)}, \\
  g_1 &= 1-M_0 . \nonumber
\end{align}
The JMN-1 model here gives the asymptotic end state of collapsing matter with vanishing radial
pressure and equal tangential pressures ($p_\theta=p_\phi$). It contains a
timelike naked singularity at $r=0$, lacks trapped surfaces, and is matched
to the Schwarzschild exterior at $r=R_{\rm b}$, with total mass
$M=M_0R_{\rm b}/2$. The bound $M_0\le4/5$ ensures subluminal sound speeds.

The shadow is the region in the observer's sky from which little or no radiation is received from an extended emission surrounding or behind the object. For null
geodesics with conserved energy $E$ and angular momentum $L$
\citep{hobson2006general}, motion in the equatorial plane obeys
\begin{equation}
  \frac{f_i}{g_i}\dot r^2 + V_{\rm eff} = 0,\quad
  V_{\rm eff} = L^2\frac{f_i}{r^2} - E^2 ,
\end{equation}
and the impact parameter $\xi=L/E$ is related to the radial turning point ($r_{\rm tp}$) by
$\xi = r_{\rm tp}/\sqrt{f_i(r_{\rm tp})}$. The radial turning point $r_{\rm tp}$ of a photon is defined as the radius where $\dot{r}=0$.

The JMN-1 interior admits no photon sphere. When matched to the Schwarzschild
exterior, a photon sphere exists at $r_{\rm ph}=3M$ provided
$R_{\rm b}\le3M$, equivalently $M_0\ge2/3$. JMN-1 therefore admits a photon
sphere only for $2/3\le M_0<4/5$. Photons with
$\xi>\xi_{\rm ph}=3\sqrt{3}M$ escape to the observer, while those with
$\xi<\xi_{\rm ph}$ are captured, so the shadow is a disc of radius $3\sqrt{3}M$ for both these analyzed ultracompact objects.

\subsection{Radiative Transfer, Image Construction and Parameter Space}

\begin{table}[b]
    \caption{\label{tab: em_models}%
    We compile below the different emissivity models which are used to construct the images of a Schwarzschild BH and JMN-1 naked singularity undergoing spherical accretion. The terms $k_{3}$ and $k_{2}$ represent arbitrary constants.}
    \renewcommand{\arraystretch}{1.5}
    \begin{ruledtabular}
        \begin{tabular}{ccc}
        \textrm{Model} & \textrm{Emissivity model} & \textrm{Reference} \\
        \colrule
        em.~1 & $j=\dfrac{\dot{m}_{\infty}\,\partial_{r}N}{16\pi^{2} r^{2} B N}$
              & \citep{kocherlakota2022} \\
        em.~2 & $j = k_{2}\, r^{-3}$ & \citep{jusufi2022} \\
        em.~3 & $j = k_{3}\, r^{-2}$ & \citep{bambi2013can} \\
        \end{tabular}
    \end{ruledtabular}
\end{table}

The intensity in ray-traced shadow images is a function only of the impact parameter in the observer plane, a result of the spherically symmetric nature of the Bondi-Michel accretion flow. Following \citep{kocherlakota2022}, the intensity integral that we are concerned with is given by Equation \ref{eq: req_int}, 

\begin{equation}
I(\xi)=
\begin{cases}
\displaystyle 
\int_{R_{\mathrm{in}}}^{D_{\rm{obs}}} 
g_{+}^{3}\, j\, \Lambda \, \mathrm{d} r,
& |\xi| < \xi_{\mathrm{ps}}, \\[8pt]

\displaystyle 
\int_{\max \left[ R_{\mathrm{in}},\, r_{\mathrm{ps}}^{+} \right]}^{D_{\rm{obs}}} 
g_{+}^{3}\, j\, \Lambda \, \mathrm{d} r,
& |\xi| = \xi_{\mathrm{ps}}, \\[8pt]

\displaystyle 
\int_{\max \left[ R_{\mathrm{in}},\, r_{\mathrm{tp}}(\xi) \right]}^{D_{\rm{obs}}}
( g_{-}^{3} + g_{+}^{3} )
j\, \Lambda \, \mathrm{d} r,
& |\xi| > \xi_{\mathrm{ps}} .
\end{cases} \label{eq: req_int}
\end{equation}\\

In Eq.~\eqref{eq: req_int}, $r_{\rm tp}$ is the radial turning point of a ray-traced null geodesic, $g=\nu_o/\nu_e$ is the redshift factor, $k^\mu$ is the photon four-momentum, $j$ is the monochromatic emission coefficient (Table~\ref{tab: em_models}), and $\Lambda=|f(r)k^t/k^r_\pm|$, where $+$ ($-$) denote motion away from (towards) the object. The integration extends to a distance $D_{\rm obs}$ much greater than the gravitational radius of the source, so that $D_{\rm obs}\to\infty$ is a valid approximation, as appropriate for sources such as Sgr~A* or M87*. Emission is truncated inside an inner cut-off radius, with $j_\nu=0$ for $r<R_{\rm in}$. Although a sharp truncation is not strictly realistic, it is a useful phenomenological model for reproducing the peak emission seen in EHT observations
\citep{kocherlakota2022}, and it allows us to quantify how the observable differences between Schwarzschild BHs and JMN-1 NSs depend on the extent of the emitting region. $R_{\rm in}$ also serves as a proxy for emission suppression near the event horizon, which may become relevant at terahertz frequencies, where high plasma densities in the innermost flow are expected to break the optically thin approximation. While not central to the present study, it may further be associated with the surface scale of exotic compact objects such as gravastars \citep{gravastar}. For JMN-1 we adopt fixed $R_{\rm in}=10^{-3}M$, allowing emission from matter arbitrarily close to the singularity. The results,
particularly the visibilities, show no significant variation between $R_{\rm in}=10^{-5}M$ and $10^{-3}M$, and change only for $R_{\rm in}=10^{-1}M$ or $1M$.

Table~\ref{tab: em_models} lists the emission coefficient models adopted in
this work. For em.~1, $N^2=f_i$ and $N^2/B^2=g_i$, following the notation of
\citet{kocherlakota2022}. For the Schwarzschild BH this gives
\begin{equation}
  j_{\rm Schw}(r)=\frac{M\dot{m}_{\infty}}{16\pi^{2}r^{3}(r-2M)},
  \label{eq: j_schw}
\end{equation}
which also holds for JMN-1 at $r>R_{\rm b}$. For $r<R_{\rm b}$,
\begin{equation}
  j_{\rm JMN\text{-}1}(r)=\frac{M\dot{m}_{\infty}R_{\rm b}^{\,k-1}}
  {16\pi^{2}(1-M_0)\,r^{3+k}},
  \label{eq: j_jmn1}
\end{equation}
where $k=M_0/[2(1-M_0)]$.


We cross-validated our Python-based code~\cite{patel_2026_zenodo} by
reproducing Fig.~1 of \citet{kocherlakota2022}, recovering the conclusions of Sec.~3 of that work. For the Schwarzschild BH, we adopt $R_{\rm in}=2$, 3, 3.6, 3.8, 5, and $6M$, including the values inferred from the M87$^\ast$ and Sgr\,A$^\ast$ ring sizes (Appendix~ \ref{app:rin}). The $R_{\rm in}$ values, the three emission models (Table~\ref{tab: em_models}), and $M_0=0.67$, 0.70, 0.75, and 0.79 for JMN-1 together define a three-dimensional parameter space over which we explore the shadow variations in subsequent sections.

\begin{figure}[t]
    \includegraphics[width=0.65\linewidth]{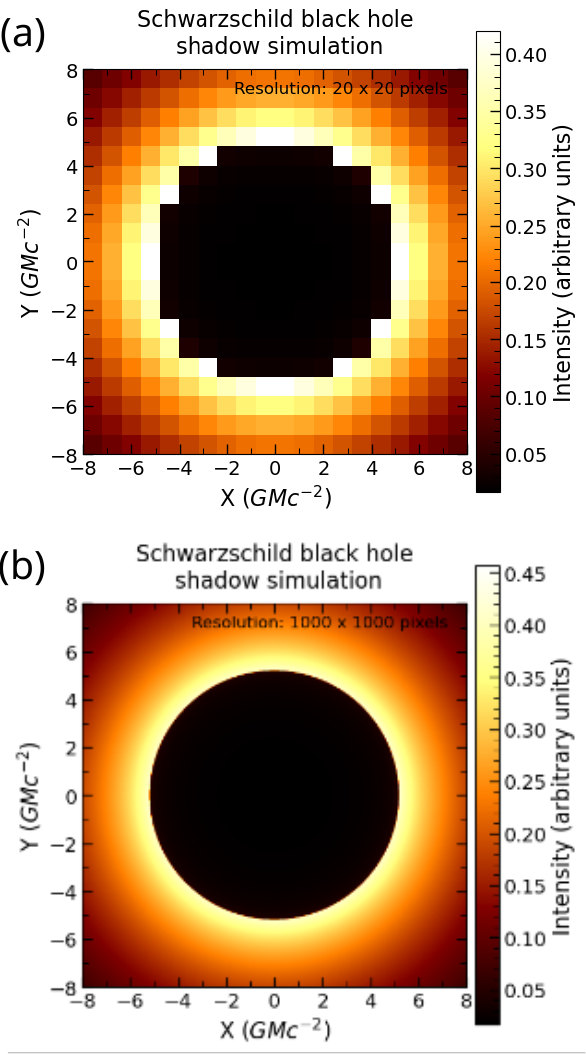}
    \caption{\label{fig:shadows_res}%
    Pixel-wise ray-traced Schwarzschild black hole shadows with
    (a) $20\times20$ pixel resolution and (b) $1000\times1000$ pixel
    resolution. Higher resolution clearly reconstructs the circular symmetry.}
\end{figure}

\section{Intensity-based shadow differentiation}

\begin{figure*}[htbp]
    \centering    \includegraphics[width=1\linewidth]{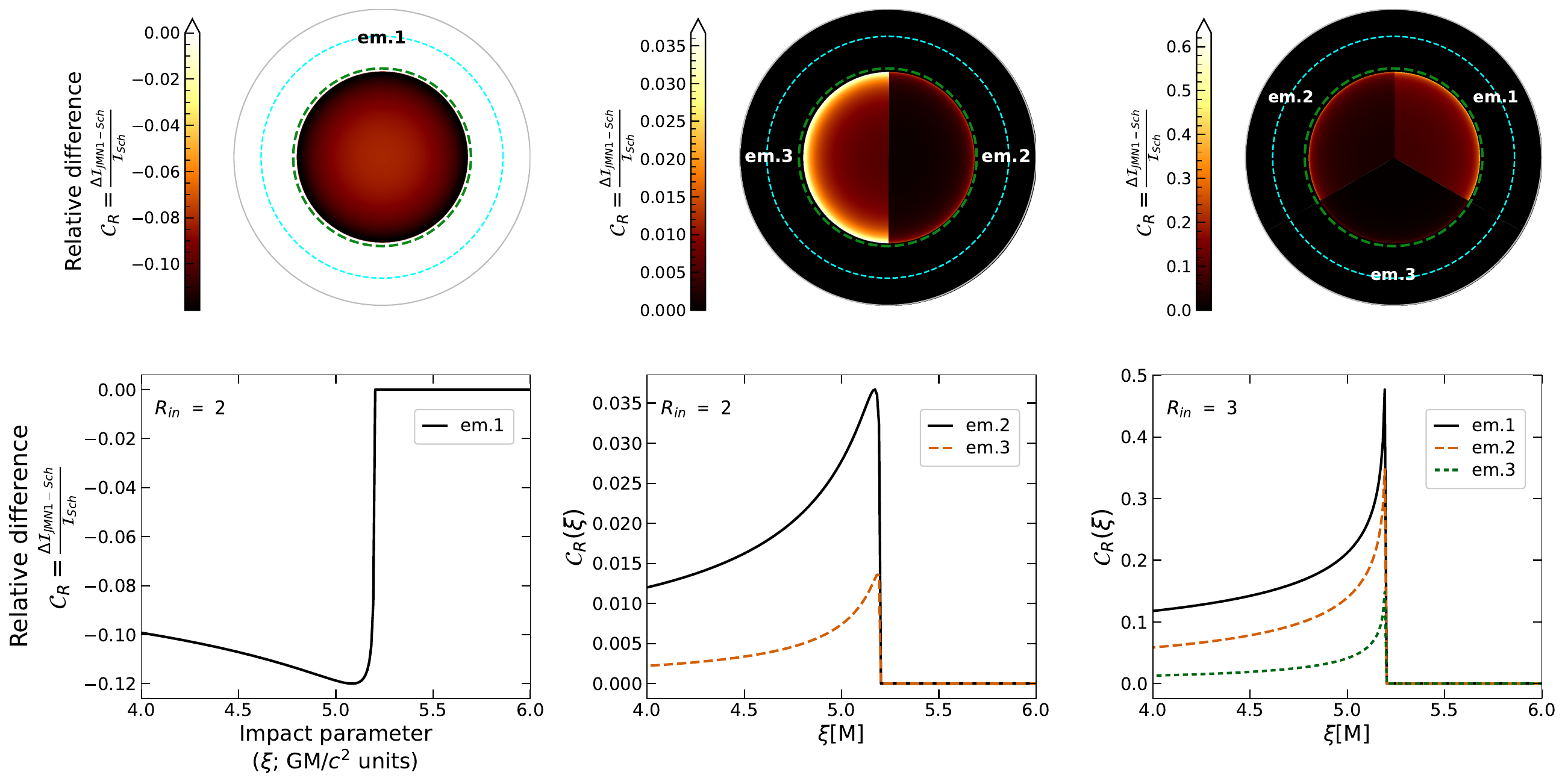}
    \caption{Relative contrast between the JMN-1 and Schwarzschild images for the emission models of Table~\ref{tab: em_models}, with $R_{\rm in}=2M$ and $3M$. Top row shows relative-contrast maps and bottom row shows the corresponding radial profiles. Circles mark the impact parameters of photons whose orbits turn at the photon sphere, $r=3M$ (black), at $r=3.8M$ (lime; $\xi\approx5.5M$, the peak of the EHT ring for M87$^*$), and at the ISCO, $r=6M$ (cyan). In all cases the contrast is confined to the shadow and sharply bounded at the photon-sphere radius. It is stronger for steeper emission profiles (larger power-law index), and except for em.1 with $R_{\rm in}=2M$, JMN-1 is the brighter image; in that case the Schwarzschild image is brighter, but the absolute contrast still peaks at the shadow edge.}
    \label{fig: diff_partI_label}
\end{figure*}

The JMN-1 NS model is regarded as the best non-spinning BH mimicker in light of
the 2022 EHT Sgr\,A$^\ast$ observations \citep[Fig.~20]{akiyama2022first},
yet, to our knowledge, a detailed intensity-based differentiation of its
shadow from that of a Schwarzschild BH has not been carried out. We quantify
this difference through the relative contrast
$C_R(\xi)=(I_{\rm JMN1}-I_{\rm Sch})/I_{\rm Sch}$, shown in
Fig.~\ref{fig: diff_partI_label} for Schwarzschild $R_{\rm in}=2$ and $3M$
(with $R_{\rm in}=10^{-3}M$ for JMN-1) and in Fig.~\ref{fig: diff_PartII_label} of the Supplemental Material for $R_{\rm in}=3.6$, 5, and $6M$. In every case, the
contrast rises towards the shadow boundary $\xi=3\sqrt{3}M$, peaks there,
and vanishes beyond it, as the Schwarzschild spacetime is the exterior metric
of JMN-1. The JMN-1 NS is uniformly brighter, except when emission extends to
the horizon ($R_{\rm in}=2M$) under em.~1 in both spacetimes, where $C_R$
turns negative yet retains the same peak and cutoff at the shadow boundary
[Fig.~\ref{fig: diff_partI_label}(a)]. As expected, the contrast strengthens with $R_{\rm in}$ and, for $j=k_nr^{-n}$, increases monotonically with $n$. Restricting to
$M_0\in[2/3,4/5]$, where JMN-1 casts a shadow, we further probe the
dependence of the contrast on $M_0$, which may help constrain this parameter
observationally (Figs.~S2 and S3). The magnitude $|C_R|$ increases
with decreasing $M_0$ for em.~2 and em.~3, but increases monotonically with
$M_0$ for em.~1, most prominently at low impact parameters. This em.~1 trend
holds even in the negative-contrast case at $R_{\rm in}=2M$, where the
Schwarzschild BH is the brighter object. We defer a detailed study of this
anomaly to future work. A direct comparison of the two intensity images, based on error-based (MSE, RMSE,
PSNR, SRE), perception-based (UIQI, SSIM, MS-SSIM, FSIM), and information-theoretic
(VIF) image-quality metrics, is presented in Section \ref{sec: image_comparison} of the Supplemental Material, where each metric is defined. The dissimilarity between the two
images increases with $R_{\rm in}$. Error-based metrics respond most strongly to this variation, whereas perceptual metrics remain largely invariant, which suggests a complementary two-step strategy in which a perceptual metric could discriminate between the spacetimes and error-based metrics could subsequently constrain $R_{\rm in}$ and the emission model. In what follows, we focus on the visibility
signatures, which, unlike the intensity distribution of the source, are directly probed by terahertz interferometric measurements.

\section{Visibility analysis and Quantum noise-limited detection}

In very long baseline interferometry (VLBI) the observable is not the source intensity but
the complex visibility which is the two-dimensional Fourier transform of the sky
brightness distribution, $I(\alpha,\beta)$.
\begin{equation}
  \mathcal{V}(u,v)=\iint I(\alpha,\beta)\,e^{-2\pi i(u\alpha+v\beta)}\,
  \mathrm{d}\alpha\,\mathrm{d}\beta,
  \label{eq: visibility}
\end{equation}
In Equation \ref{eq: visibility}, $(u,v)$ are the baseline coordinates in units of the observing wavelength.
For the spherically symmetric Bondi--Michel flow considered here, the intensity distributions are
radially symmetric, so $\mathcal{V}$ depends only on the baseline length
$L=\sqrt{u^2+v^2}$ and Eq.~\eqref{eq: visibility} reduces to the real-valued
zeroth-order Hankel transform \citep{osgood2002lecture}
\begin{equation}
  \mathcal{V}(L)=2\pi\int_0^\infty I(\rho)\,J_0(2\pi L\rho)\,\rho\,\mathrm{d}\rho,
  \label{eq: visibility_hankel_main}
\end{equation}
with $\rho=\sqrt{\alpha^2+\beta^2}$, the angular radius on the sky. A similar approach has been previously used in the literature by \citet{chen2024interferometric} to simulate the
interferometric signatures of black holes with multiple photon spheres.

An ideal Bondi--Michel flow extends to infinity, with an intensity that falls only
as a power law in the impact parameter $\xi$. We therefore truncate the image with
a raised-cosine taper (Eq.~\ref{eq:taper}). The tapering function leaves the intensity unchanged for $\xi\le\xi_0$, rolls it off smoothly for $\xi_0<\xi<\xi_1$, and sets it to zero for $\xi> \xi_1$, mimicking the dark sky background. The smooth roll-off avoids the ringing in the visibilities that a
sharp cut-off would cause, and choosing $\xi_0>\xi_c=3\sqrt{3}\,M$ keeps the taper
in the region where the Schwarzschild and JMN-1 images coincide. The taper edges $(\xi_0, \xi_1)$ are fixed by the compact size of Sgr~A$^*$
measured by the EHT on its shortest inter-site baseline (SMT--LMT,
$\sim 0.6$--$1.5\,\mathrm{G}\lambda$). We adopt the $60\,\mu$as size seen on
baselines shorter than $2\,\mathrm{G}\lambda$ \citep{EHT2022_SgrA_III} to obtain
the solution set for the $(\xi_0, \xi_1)$ parameters. Appendix~\ref{app:taper}
details the adopted procedure and the resulting $(\xi_0, \xi_1)$ solutions, all of
which reproduce the $60\,\mu$as size at the short inter-site SMT--LMT baseline.

\begin{figure*}
\centering
\includegraphics[width=0.9\textwidth]{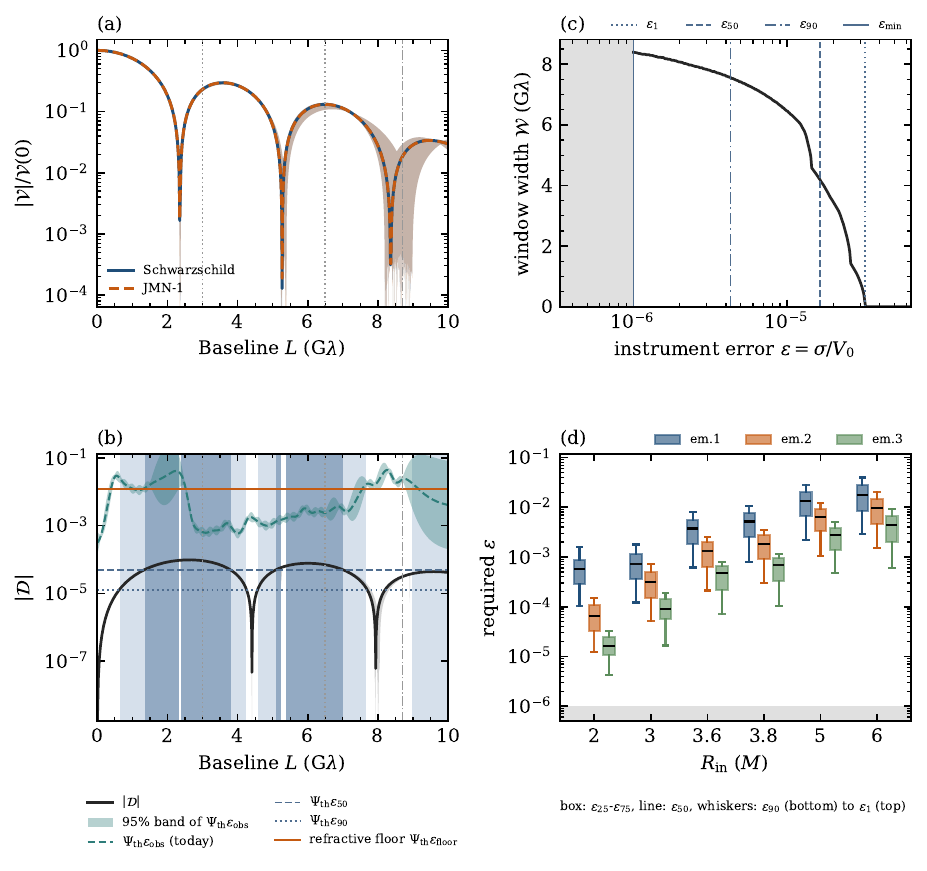}
\caption{Instrument error needed to distinguish a Schwarzschild black hole (BH)
from a JMN-1 naked singularity (NS) in the visibilities of Sgr~A$^*$. Panels
(a)--(c) show a representative case (em.3, $R_{\mathrm{in}}=2M$, source size
$60\,\mu$as) and panel (d) shows all 18 cases. The errors are fractions of the total
flux, $\epsilon=\sigma/V_0$ ($V_0=2.2$\,Jy). (a)~Normalized amplitudes
$\mathcal{A}_X(L)=|\mathcal{V}_X(L)|/\mathcal{V}_X(0)$ of the BH (blue, solid) and
NS (orange, dashed); bands span all taper pairs $(\xi_0,\xi_1)$ reproducing the
$60\,\mu$as size, and bold lines show a representative pair. (b)~Difference
$|\mathcal{D}|$ (black, grey band over tapers) against the thresholds
$\Psi_{\mathrm{th}}\epsilon$ ($\Psi_{\mathrm{th}}=3$), namely the 2017 EHT
statistical uncertainty $\Psi_{\mathrm{th}}\epsilon_{\mathrm{obs}}$ (teal,
Gaussian-process fit with its 95\% prediction band, which widens at baselines
lacking data), two improved instrument sensitivities,
$\Psi_{\mathrm{th}}\epsilon_{50}$ and $\Psi_{\mathrm{th}}\epsilon_{90}$ (slate-blue
dashed and dotted), and the refractive-scattering floor
$\Psi_{\mathrm{th}}\mathcal{\epsilon}_{\mathrm{floor}}$ (orange,
$\mathcal{\epsilon}_{\mathrm{floor}}=4\times10^{-3}$). Vertical shading marks the detection
windows at $\epsilon_{90}$ (light) and $\epsilon_{50}$ (dark). (c)~Window width
$\mathcal{W}(\epsilon)$, with $\epsilon_1$ (dotted, first detection),
$\epsilon_{50}$ and $\epsilon_{90}$ (dashed and dash-dotted, 50\% and 90\% of
$\mathcal{W}_{\mathrm{best}}$), and the assumed instrument limit
$\epsilon_{\min}=10^{-6}$ (solid; grey shading marks errors below this limit, here
and in (d)). (d)~Required error for all cases, grouped by $R_{\mathrm{in}}$ and
coloured by emissivity model (em.1 blue, em.2 orange, em.3 green); boxes span the
$\epsilon_{25}$--$\epsilon_{75}$ quartile range, black lines mark $\epsilon_{50}$,
and whiskers run from $\epsilon_{90}$ to $\epsilon_1$. In (a) and (b), dotted lines
mark the observed visibility minima of Sgr~A$^*$ at $\approx3.0$ and
$6.5\,\mathrm{G}\lambda$~\cite{EHT2022_SgrA_III}, and the dash-dotted line denotes the
longest EHT baseline, $\approx8.7\,\mathrm{G}\lambda$~\cite{EHT2022_SgrA_III}. In
the representative case, both the 2017 uncertainties and the refractive floor
exceed the required uncertainty on every baseline, so the instrument error must be
lowered and the floor averaged down over many epochs. Across emission models the
requirement varies by about three orders of magnitude.}
\label{fig: visibility_signature}
\end{figure*}

Figure~\ref{fig: visibility_signature}(a) shows the normalized visibility amplitude
\begin{equation}
\mathcal{A}_X(L)=\frac{|\mathcal{V}_X(L)|}{\mathcal{V}_X(0)},\qquad
X\in\{\mathrm{Sch},\,\mathrm{JMN\text{-}1}\},
\label{eq:amp}
\end{equation}
where $L$ is the projected baseline length in units of the observing wavelength,
$\mathcal{V}_X(L)$ is the visibility of the tapered image of spacetime $X$, and
$\mathcal{V}_X(0)\equiv V_0$ is its intrasite or zero-baseline visibility amplitude, i.e.\ the total flux of that image.
Normalizing by $V_0$ prevents the requirement for absolute flux calibration. $\mathcal{A}_X(L)$, therefore gives the
fraction of the total flux seen on baseline $L$, which implies $\mathcal{A}_X(0)=1$. Each
$(\xi_0,\xi_1)$ pair of the solution set gives one visibility amplitude v/s baseline curve. The bold line in Figure~\ref{fig: visibility_signature}(a) highlights a representative pair, while the shaded bands represent variation across all such pairs  (Appendix~\ref{app:taper}).

The two spacetimes differ by
$\mathcal{D}(L)=\mathcal{A}_{\mathrm{JMN\text{-}1}}(L)-\mathcal{A}_{\mathrm{Sch}}(L)$. Let the normalized uncertainty be $\epsilon = \sigma/V_0$. To make the result hold for every allowed taper, we use the guaranteed difference
$\mathcal{D}_{\mathrm{lo}}(L)$, which is the least $|\mathcal{D}(L)|$ over the
solution set, set to zero where the pairs disagree on its sign. We use the three-sigma criteria to define required signal-to-noise ratio (SNR) (i.e. $\Psi_{\mathrm{th}}=3$), to define an observationally significant distinction between the two singularities. Hence, the difference is detected if $\Psi(L) = |\mathcal{D}(L)|/ \epsilon = |\mathcal{D}(L)|\,V_0/\sigma\geq\Psi_{\mathrm{th}}$.  Equation \ref{eq: epsilon_req} now gives the minimum required sensitivity, $\epsilon_{\mathrm{req}}$, for achieving the prescribed SNR. 

\begin{equation}
\epsilon\leq\epsilon_{\mathrm{req}}(L)=\mathcal{D}_{\mathrm{lo}}(L)/\Psi_{\mathrm{th}}. 
\label{eq: epsilon_req}
\end{equation}

Figure~\ref{fig: visibility_signature}(b) compares $|\mathcal{D}|$ with the
detection threshold $\Psi_{\mathrm{th}}\epsilon$ for three noise levels. The first
is set by the statistical (thermal-noise) uncertainties of the 2017 April 7 EHT observations,
$\epsilon_{\mathrm{obs}}(L)=\sigma_{\mathrm{obs}}(L)/V_0$, where
$\sigma_{\mathrm{obs}}$ is the smallest error attained in each
$0.2\,\mathrm{G}\lambda$ baseline bin. The data points are interpolated by a Gaussian process into a
continuous function of baseline. The depicted teal band marks the 95\% uncertainty of this
prediction. The other two correspond to improved instrument sensitivities with errors $\epsilon_{50}$ and $\epsilon_{90}$, defined below. The
panel also shows the refractive-scattering floor $\Psi_{\mathrm{th}}\mathcal{\epsilon}_{\mathrm{floor}}$, with
$\mathcal{\epsilon}_{\mathrm{floor}}=4\times10^{-3}$ \citep{EHT2022_SgrA_III} for Sgr~A$^*$, below which no instrument can detect a difference in a single epoch. Shading marks the detection windows, the baselines where $\mathcal{D}_{\mathrm{lo}}$ exceeds the threshold.

A smaller $\epsilon$ opens more baselines. The detection window at $\epsilon$ is
$\{L:\epsilon_{\mathrm{req}}(L)\geq\epsilon\}$, and
$\mathcal{W}(\epsilon)$ in Figure~\ref{fig: visibility_signature}(b) is its total width at sensitivity of $\epsilon$. $\mathcal{W}$ increases with decrease in $\epsilon$ and saturates at $\mathcal{W}_{\mathrm{best}}=\mathcal{W}(\epsilon_{\min})$. We set $\epsilon_{\min}=10^{-6}$ as the assumed minimum achievable instrument noise. Three numbers $\epsilon_1$, $\epsilon_{50}$ and $\epsilon_{90}$ describe this curve. Here $\epsilon_1=\max_L\epsilon_{\mathrm{req}}(L)
=\mathcal{D}_{\mathrm{lo,max}}/\Psi_{\mathrm{th}}$, at which the first window opens on the single baseline $L_{\mathrm{best}}$ where the guaranteed difference is largest. $\epsilon_{50}$ and $\epsilon_{90}$ are the largest normalized errors whose windows cover 50\% and 90\% of $\mathcal{W}_{\mathrm{best}}$. Figure~\ref{fig: visibility_signature}(d) repeats this for all 18 cases (three emissivity models each with six $R_{\mathrm{in}}$ values), where each box spans $\epsilon_{25}$--$\epsilon_{75}$ quartile and the line marks $\epsilon_{50}$ value. The whiskers represent the $\epsilon_{90}$ to $\epsilon_1$ range. 

We find that the requirement depends strongly on the emission model (Table~\ref{tab: cases} of
the Supplemental Material). Across the 18 cases, the largest guaranteed
difference spans three orders of magnitude, from
$\mathcal{D}_{\mathrm{lo,max}}=9.7\times10^{-5}$ (em.3, $R_{\mathrm{in}}=2M$) to
$1.2\times10^{-1}$ (em.1, $R_{\mathrm{in}}=6M$). For every emissivity model it
increases with $R_{\mathrm{in}}$, and at fixed $R_{\mathrm{in}}$ it is largest for
em.1 and smallest for em.3, increasing with the steepness of the emissivity
profile. Accordingly, $\epsilon_1$ ranges from $3.2\times10^{-5}$ to
$4.0\times10^{-2}$, and $\epsilon_{50}$ ($\epsilon_{90}$) from $1.6\times10^{-5}$
($4.3\times10^{-6}$) to $1.8\times10^{-2}$ ($2.9\times10^{-3}$). The guaranteed
difference peaks at $L_{\mathrm{best}}\simeq2.5$--$2.6\,\mathrm{G}\lambda$ for
$R_{\mathrm{in}}\leq3.8M$ and shifts to $5.2$--$5.6\,\mathrm{G}\lambda$ for most
cases with $R_{\mathrm{in}}\geq5M$.

A subset of baselines remains effective for every model. The $\epsilon_{50}$
windows of all 18 cases overlap on $1.43$--$2.34$, $2.55$--$3.12$, and
$5.94$--$6.23\,\mathrm{G}\lambda$ ($1.8\,\mathrm{G}\lambda$ in total), and an error
$\epsilon=1.6\times10^{-5}$ ($\sigma=36\,\mu$Jy for $V_0=2.2$\,Jy) opens all of
them. The single best baseline for all cases at once is
$L_\star=2.64\,\mathrm{G}\lambda$, where $\epsilon_\star=3.2\times10^{-5}$
($\sigma_\star=71\,\mu$Jy) detects the difference in every case (Fig.~\ref{fig: windows} of the
Supplemental Material). At $L_\star=2.64\,\mathrm{G}\lambda$, the Gaussian-process estimate of the 2017
statistical uncertainty is $\sigma_{\mathrm{obs}}\simeq2.0$\,mJy, so a roughly
28-fold reduction is needed to ensure an emission-model-agnostic distinction
between the spacetimes.

The source noise imposes a second, independent limit. A single epoch reaches
$\Psi_{\mathrm{th}}$ only if
$\mathcal{D}_{\mathrm{lo,max}}>\Psi_{\mathrm{th}}\mathcal{\epsilon}_{\mathrm{floor}}
=1.2\times10^{-2}$. This holds in 8 of the 18 cases, all with extended emission
($R_{\mathrm{in}}\geq5M$, and em.1 with $R_{\mathrm{in}}=3.6$ and $3.8M$), and in
these cases the 2017 statistical uncertainties already satisfy
$\epsilon_{\mathrm{obs}}(L_{\mathrm{best}})<\epsilon_1$. In the remaining ten, both
limits must be overcome. The instrument error must fall by a factor of $1.1$--$28$,
and the refractive floor must be averaged down over $N_{\mathrm{ep}}\geq(\Psi_{\mathrm{th}}\mathcal{\epsilon}_{\mathrm{floor}}/
\mathcal{D}_{\mathrm{lo,max}})^2$ independent epochs, from $\approx1.3$ to
$1.5\times10^{4}$.

Therefore, whether these signatures can be observed greatly depends on the array's sensitivity. Combining the radiometer equation with azimuthal averaging of $N_b$ baselines of equal length (Appendix~\ref{app:sensitivity}), we find that the error
$\epsilon_{\min}=10^{-6}$ ($\sigma=2.2\,\mu$Jy), at which the detection window is widest in every case, requires a coherent integration time $t_{\rm int}\gtrsim220$~s for an array of near-quantum-limited 230~GHz receivers ($T_{\rm sys}\approx27.6$~K, $\Delta\nu\approx10$~GHz) with at most five baselines and $A_{\rm eff}\le2A_{\rm ALMA}$ (Table~\ref{tab:sefd_requirements}).
This lies far beyond the coherence time allowed by tropospheric turbulence on Earth, but is attainable in principle in space, where atmospheric phase fluctuations are absent. Section~\ref{sec: sm_architecture} of the Supplemental Material outlines a candidate architecture that uses continuous multi-beam phase referencing to
sustain coherence over such times, although this remains to be verified experimentally. Such a space interferometer would reach the instrument error required to distinguish the Schwarzschild BH and JMN-1 NS spacetimes; for the most compact emission models, the refractive floor must also be averaged over many epochs.

\section{Conclusion and Future Work}

We have quantified, in the visibility domain, the difference between a
Schwarzschild BH and a JMN-1 NS embedded in optically thin spherical accretion, and translated it into instrument requirements for Sgr~A$^*$. Requiring the difference to hold for every image truncation consistent with the observed source size, we find that the required per-visibility error spans three orders of magnitude across emission models. Independently of the emission model, the two spacetimes become separable on $1.43$--$2.34$, $2.55$--$3.12$, and $5.94$--$6.23\,\mathrm{G}\lambda$ at $\sigma\simeq36\,\mu$Jy, and on the most favourable baseline, $L_\star=2.64\,\mathrm{G}\lambda$, at $\sigma\simeq71\,\mu$Jy,
about 28 times below the 2017 EHT statistical uncertainty. When the emission begins far from the horizon, the difference is already detectable at the 2017 sensitivity and exceeds the refractive-scattering floor in a single epoch, whereas emission reaching closer to the horizon requires both a lower instrument error and averaging over many epochs. Opening the widest detection window demands coherent integrations of several minutes, which favours a near-quantum-limited space interferometer.

These conclusions rest on idealized assumptions, namely spherical, optically thin emission, statistical uncertainties without calibration systematics, and baselines characterized by length alone. A natural next step is to adopt a radiatively inefficient accretion flow with full radiative transfer and to pass both spacetimes through synthetic observations with realistic array coverage, as recently done for other naked singularities~\cite{deliyski2025}, thereby testing whether the predicted baselines and requirements survive realistic reconstruction. General-relativistic magnetohydrodynamic simulations of accretion onto JMN-1 would then establish whether the signatures identified here persist for turbulent, non-spherical flows, and thus whether they can serve as a robust test of cosmic censorship.

\section*{Data Availability}
The complete Python codebase developed in this work is published as a Zenodo record~\cite{patel_2026_zenodo} and in the associated GitHub repository \href{https://github.com/vpatel2000/Quantum-Limited-Thz-VLBI.git}{Quantum-Limited-Thz-VLBI} \cite{patel2026code}. Both also contain the derived data products, namely the per-visibility error curve obtained from the publicly available EHT 2017 data~\cite{EHT2022_SgrA_II} and the results for all model cases, which the code can regenerate.


\section{Acknowledgments}

\begin{acknowledgments}
This study is part of an advanced research and technology development program at
the Space Applications Centre (SAC), Indian Space Research Organisation, in collaboration with the International Centre for Space and Cosmology (ICSC),
Ahmedabad University. The project is funded by SAC under TDP code \textit{202506021}. We thank Dr.~Rashmi Sharma, Deputy Director, Earth and Planetary Sciences Area, SAC, and Dr.~Ch.~V.~N.~Rao, Director, SAC, for their constant motivation and support. We also thank Koushiki Malda and our colleagues at ICSC for helpful comments and discussions, and Lakshay Dhiman for initial work on benchmarking image-comparison metrics during an ISRO internship. Large language model tools, namely Anthropic's Claude Opus 5.5 and Google's Antigravity (powered by Gemini models), assisted with code development, brainstorming, and language editing. We verified the results and take full responsibility for them. This work also uses the Event Horizon Telescope 2017 Sgr~A$^*$ calibrated data products~\cite{EHT2022_SgrA_II}, obtained from the
CyVerse Data Commons (DOI: 10.25739/m140-ct59).
\end{acknowledgments}

\section*{Appendix}        
\appendix
\setcounter{secnumdepth}{2}  

\section{\label{app:rin} $R_{\rm in}$ parameter space}

Following the analysis of \citet{kocherlakota2022} for M87$^\ast$, we also
constrain the inner emission cut-off radius for Sgr\,A$^\ast$, which, to our
knowledge, has not been done within this framework. The EHT ring sizes give
$\xi_{\rm peak}/M=\alpha/2\approx5.5$ for M87$^\ast$ and $\approx5.4$ for
Sgr\,A$^\ast$, using $\alpha=10.8^{+1.3}_{-2.1}$ \citep{event2022first}.
Within the Schwarzschild emission framework, this requires
$R_{\rm in}\approx3.77M$ and $3.6M$, respectively. Despite the simplicity of
the accretion model, these results remain broadly consistent with the 2019
M87$^\ast$ and 2022 Sgr\,A$^\ast$ EHT observations, supporting the conclusion of dominant emission from radii close to the photon sphere. The adopted values
$R_{\rm in}=2$, 3, 3.6, 3.8, 5, and $6M$ therefore span the horizon ($2M$) and
photon sphere ($3M$) to the ISCO ($6M$), and include the Sgr\,A$^\ast$ and
M87$^\ast$ constraints.

\section{Calibration of the outer taper}
\label{app:taper}

\subsection{Outer taper and visibility}
\label{app:taper_def}

The ray-traced intensity $I(\xi)$, with the impact parameter $\xi$ in units of $M$, is multiplied
in the image plane by the raised-cosine taper
\begin{equation}
T(\xi) =
\begin{cases}
1, & \xi \le \xi_0,\\
\tfrac12\!\left[1+\cos\!\left(\pi\,\dfrac{\xi-\xi_0}{\xi_1-\xi_0}\right)\right], & \xi_0<\xi<\xi_1,\\
0, & \xi \ge \xi_1,
\end{cases}
\label{eq:taper}
\end{equation}
with $\xi_{\rm ph} < \xi_0 < \xi_1$. The taper acts on the image rather than on the emissivity,
so every ray is still integrated along its full path. To compare with interferometric data, the
impact parameter is converted to an angle on the sky, $\beta = \theta_g\,\xi$, where
$\theta_g = GM/c^2D$ is the angle subtended by one gravitational radius at distance $D$. For a
circularly symmetric image the complex visibility on a baseline of length $k$ (in wavelengths)
reduces to the zeroth-order Hankel transform (Equation \ref{eq: visibility_hankel_main}) with $\rho$ in radians. Since the two images coincide for $\xi > \xi_{\rm ph}$ and $T=1$ for
$\xi<\xi_0$, the difference $V_{\rm JMN}-V_{\rm Sch}$ involves only $\xi<\xi_{\rm ph}$ and is
independent of $(\xi_0,\xi_1)$. However, quantities that involve a single model, such
as the normalized amplitude $|V(u)|/V(0)$ and the positions of the visibility nulls, do depend on
the taper, which is why $(\xi_0,\xi_1)$ must be fixed by observation
(Appendices~\ref{app:theta_eq} and \ref{app:xi_curves}).

\subsection{\label{app:theta_eq} Equivalent Gaussian size}
A circular Gaussian of compact flux $F_{\rm cpt}$ and FWHM $\theta$ has
$|V(u)| = F_{\rm cpt}\exp[-\pi^2\theta^2u^2/(4\ln2)]$, so that
\begin{equation}
\theta_{\rm eq} = \frac{2\sqrt{\ln2}}{\pi u}\sqrt{\ln\frac{V(0)}{|V(u)|}} .
\label{eq:theta_eq}
\end{equation}
The EHT applies this relation to the SMT--LMT baseline ($\sim0.6$--$1.5\,\mathrm{G}\lambda$) with
the total flux in place of the unmeasured compact flux, which turns it into an upper limit on the
compact size \citep[cf.][Eq.~8]{EHT2022_SgrA_II}. Post-truncation, $V(0)$ can be approximated as the compact flux itself and
Eq.~(\ref{eq:theta_eq}) holds as an equality, which we evaluate at $u = 1\,\mathrm{G}\lambda$. We
match it to the size used in the EHT data processing, which is $60\,\mu$as \citep{EHT2022_SgrA_III}, and to
the upper limit, $87\,\mu$as \citep{EHT2022_SgrA_II} to constrain $(\xi_0,\xi_1)$ values in each of these scenarios.

\subsection{Admissible taper edges\label{app:xi_curves}}
\begin{figure}
\centering
\includegraphics[width=\linewidth]{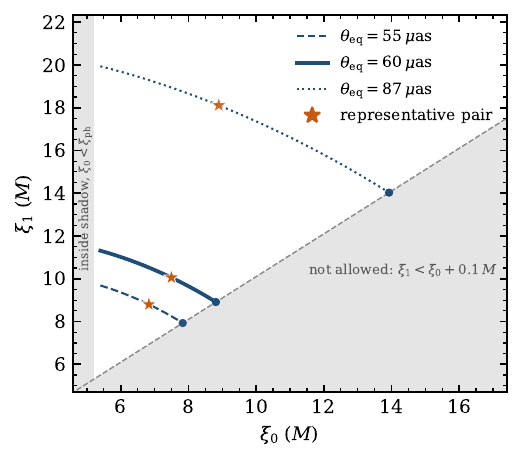}
\caption{Taper edges $(\xi_0,\xi_1)$ reproducing the EHT short-baseline size of Sgr~A$^*$:
$60\,\mu$as (solid) and $87\,\mu$as (dashed). Dots mark $\xi_0^{\max}$ and stars the pair used in
the main text. Shaded: excluded regions ($\xi_0<\xi_{\rm ph}$, $\xi_1<\xi_0+\Delta_{\min}$, and
$\xi_1$ beyond the grid).}
\label{fig:xi_curves}
\end{figure}

One measured or target source size, constrains $(\xi_0,\xi_1)$. Hence, for each target
$\theta_\star\in\{60,87\}\,\mu$as the admissible $(\xi_0,\xi_1)$ values form a curve. At fixed $\xi_0$,
$\theta_{\rm eq}$ increases monotonically with $\xi_1$, so
$\theta_{\rm eq}(\xi_0,\xi_1)=\theta_\star$ has at most one root, which we find with Brent's method
on $[\xi_0+\Delta_{\min},\,\xi_{\max}]$ ($\Delta_{\min}=0.1\,M$, $\xi_{\max}$ must be within the ray-traced
grid). The curve runs from $\xi_0=\xi_{\rm ph}+0.2\,M$ to the end point $\xi_0^{\max}$, defined by
$\theta_{\rm eq}(\xi_0^{\max},\xi_0^{\max}+\Delta_{\min})=\theta_\star$
(Fig.~\ref{fig:xi_curves}).Along each curve, increasing $\xi_0$ requires a smaller $\xi_1$, i.e.\ a narrower roll-off, to keep the size fixed. 

\section{Sensitivity requirements}
\label{app:sensitivity}

For a baseline between stations $i$ and $j$, the thermal uncertainty of a complex
visibility follows the radiometer equation \cite{thompson2017}. Because the
visibility of a spherically symmetric source depends only on baseline length,
azimuthally averaging $N_b$ independent baselines of equal length into one radial
bin reduces this uncertainty by $\sqrt{N_b}$ \cite{nrao2013sensitivity}:
\begin{equation}
  \sigma_{\mathcal{V},\rm eff}=\frac{1}{\eta_c}
  \sqrt{\frac{\mathrm{SEFD}_i\,\mathrm{SEFD}_j}
  {2N_b\,n_{\rm pol}\,\Delta\nu\,t_{\rm int}}},
  \label{eq:radial_noise}
\end{equation}
where $\eta_c$ is the correlator efficiency, $n_{\rm pol}$ the number of
polarization products ($n_{\rm pol}=2$ for Stokes I, Q, U, or V images),
$\Delta\nu$ the bandwidth, and $t_{\rm int}$ the coherent integration time. $N_b=n_{\rm pol}=1$ recovers the single-baseline radiometer equation. The system
equivalent flux density,
\begin{equation}
  \mathrm{SEFD}=\frac{2k_BT_{\rm sys}}{A_{\rm eff}},
  \label{eq:sefd}
\end{equation}
is the flux density of a point source in the main beam that doubles the system
noise power of a station. Here $A_{\rm eff}$ is the effective collecting area and
$T_{\rm sys}=T_{\rm rec}+T_{\rm atm}+T_{\rm spill}+T_{\rm sky}$ the sum of the
receiver, atmospheric, spillover, and sky contributions \cite{thompson2017}.

\begin{table}[htbp]
\caption{\label{tab:sefd_requirements}%
Shortest coherent integration time $t_{\rm int}$ needed to reach a visibility
error $\epsilon_{\min}=10^{-6}$ ($\sigma=2.2~\mu$Jy for $V_0=2.2$~Jy) with $N_b$
identical baselines, for stations of the largest allowed effective collecting area,
$A_{\rm eff}=2A_{\rm ALMA}$ ($A_{\rm ALMA}\approx2846~\mathrm{m^2}$). using Eq.~\eqref{eq:sefd} with $T_{\rm sys}=27.6$~K, this area gives the smallest achievable SEFD ($\mathrm{SEFD}_{\min}$). The derived $\mathrm{SEFD}_{\min}$ in Eq.~\eqref{eq:radial_noise} yields $t_{\rm int}$. Given $\Delta\nu=10$~GHz,
$n_{\rm pol}=2$, and $\eta_c=0.92$; values of $t_{\rm int}$ are rounded up to 5~s.}
\begin{ruledtabular}
\begin{tabular}{cccc}
$N_b$ & $A_{\rm eff}/A_{\rm ALMA}$ & $\mathrm{SEFD}_{\min}$ (Jy) & $t_{\rm int}$ (s) \\
\colrule
2 & 2.00 & 13.4 & 550 \\
3 & 2.00 & 13.4 & 365 \\
4 & 2.00 & 13.4 & 275 \\
5 & 2.00 & 13.4 & 220 \\
\end{tabular}
\end{ruledtabular}
\end{table}

We assume a homogeneous array of dual-polarization stations ($n_{\rm pol}=2$) with
$\eta_c=0.92$~\cite{nrao2013sensitivity} and $\Delta\nu\approx10$~GHz, set by the
2--12~GHz IF band of near-quantum-limited 230~GHz endfire SIS
mixers~\citep{endfireSIS}. A space-based receiver with a millikelvin-cooled
parametric IF amplifier could reach $T_{\rm sys}\approx27.6$~K, as outlined for the
candidate architecture in Section~\ref{sec: sm_architecture} of the Supplemental Material. We limit the array to $N_b\le5$ identical baselines and stations of effective area $A_{\rm eff}\le2A_{\rm ALMA}$, where $A_{\rm ALMA}\approx2846~\mathrm{m^2}$ is the effective area of the 37 12-m antennas of the phased ALMA array at the Band~6 aperture efficiency of 0.68~\citep{almaArea}. At the largest allowed area,
Eq.~\eqref{eq:sefd} gives the smallest achievable SEFD, 13.4~Jy, and inverting Eq.~\eqref{eq:radial_noise} with $\sigma_{\mathcal{V},\rm eff}=\epsilon V_0$ then
yields the shortest coherent integration time for each $N_b$, listed in Table~\ref{tab:sefd_requirements}.

Reaching $\epsilon_{\min}$ under these limits therefore requires
$t_{\rm int}\gtrsim220$~s even at the largest $N_b=5$, rising to $\gtrsim275$~s for
$N_b=4$. Since $t_{\rm int}\propto\epsilon^{-2}$ at fixed SEFD and $N_b$, the less
demanding sensitivities of the main text need much shorter times. For example, the
common $\epsilon_{90}$ window ($\epsilon=4.3\times10^{-6}$) requires
$t_{\rm int}\gtrsim12$~s and the common $\epsilon_{50}$ window
($\epsilon=1.6\times10^{-5}$) $t_{\rm int}\gtrsim1$~s, both at $N_b=5$.

\bibliographystyle{apsrev4-2}
\bibliography{main}

@book{hobson2006general,
  title={General relativity: an introduction for physicists},
  author={Hobson, Michael Paul and Efstathiou, George P and Lasenby, Anthony N},
  year={2006},
  publisher={Cambridge university press}
}

@article{joshi2011equilibrium,
  title={Equilibrium configurations from gravitational collapse},
  author={Joshi, Pankaj S and Malafarina, Daniele and Narayan, Ramesh},
  journal={Classical and Quantum Gravity},
  volume={28},
  number={23},
  pages={235018},
  year={2011},
  publisher={IOP Publishing},
  doi= {10.1088/0264-9381/28/23/235018}
}

@article{joshi2014distinguishing,
  title={Distinguishing black holes from naked singularities through their accretion disc properties},
  author={Joshi, Pankaj S and Malafarina, Daniele and Narayan, Ramesh},
  journal={Classical and Quantum Gravity},
  volume={31},
  number={1},
  pages={015002},
  year={2014},
  publisher={IOP Publishing},
doi = {10.1088/0264-9381/31/1/015002}
}

@article{bambi2013can,
  title={Can the supermassive objects at the centers of galaxies be traversable wormholes? The first test<? format?> of strong gravity for mm/sub-mm very long baseline interferometry facilities},
  author={Bambi, Cosimo},
  journal={Physical Review D—Particles, Fields, Gravitation, and Cosmology},
  volume={87},
  number={10},
  pages={107501},
  year={2013},
  publisher={APS},
doi = {https://doi.org/10.1103/PhysRevD.87.107501}
}

@article{kaur2021comparing,
  title={Comparing shadows of blackhole and naked singularity},
  author={Kaur, Kanwar Preet and Joshi, Pankaj S and Dey, Dipanjan and Joshi, Ashok B and Desai, Rucha P},
  journal={arXiv preprint arXiv:2106.13175},
  year={2021},
doi = {https://doi.org/10.48550/arXiv.2106.13175}
}

@article{akiyama2022first,
  title={First Sagittarius A* event horizon telescope results. VI. Testing the black hole metric},
  author={Akiyama, Kazunori and Alberdi, Antxon and Alef, Walter and Algaba, Juan Carlos and Anantua, Richard and Asada, Keiichi and Azulay, Rebecca and Bach, Uwe and Baczko, Anne-Kathrin and Ball, David and others},
  journal={The Astrophysical Journal Letters},
  volume={930},
  number={2},
  pages={L17},
  year={2022},
  publisher={IOP Publishing},
doi = {10.3847/2041-8213/ac6756F}
}

@article{shaikh2019shadows,
  title={Shadows of spherically symmetric black holes and naked singularities},
  author={Shaikh, Rajibul and Kocherlakota, Prashant and Narayan, Ramesh and Joshi, Pankaj S},
  journal={Monthly Notices of the Royal Astronomical Society},
  volume={482},
  number={1},
  pages={52--64},
  year={2019},
  publisher={Oxford University Press},
doi = {https://doi.org/10.1093/mnras/sty2624}
}

@article{shaikh2019can,
  title={Can we distinguish black holes from naked singularities by the images of their accretion disks?},
  author={Shaikh, Rajibul and Joshi, Pankaj S},
  journal={Journal of Cosmology and Astroparticle Physics},
  volume={2019},
  number={10},
  pages={064},
  year={2019},
  publisher={IOP Publishing},
  doi = {10.1088/1475-7516/2019/10/064}
}

@article{narayan2019shadow,
  title={The shadow of a spherically accreting black hole},
  author={Narayan, Ramesh and Johnson, Michael D and Gammie, Charles F},
  journal={The Astrophysical Journal Letters},
  volume={885},
  number={2},
  pages={L33},
  year={2019},
  publisher={IOP Publishing},
  doi = {10.3847/2041-8213/ab518c}
}

@article{kocherlakota2022,
  title={Distinguishing gravitational and emission physics in black hole imaging: spherical symmetry},
  author={Kocherlakota, Prashant and Rezzolla, Luciano},
  journal={Monthly Notices of the Royal Astronomical Society},
  volume={513},
  number={1},
  pages={1229--1243},
  year={2022},
  publisher={Oxford University Press}
}

@article{jusufi2022,
  title={Constraining wormhole geometries using the orbit of S2 star and the Event Horizon Telescope},
  author={Jusufi, Kimet and Kumar, Saurabh and Azreg-A{\"\i}nou, Mustapha and Jamil, Mubasher and Wu, Qiang and Bambi, Cosimo},
  journal={The European Physical Journal C},
  volume={82},
  number={7},
  pages={633},
  year={2022},
  publisher={Springer}
}

@article{event2022first,
  title={First Sagittarius A* Event Horizon Telescope results. IV. Variability, morphology, and black hole mass},
  author={Akiyama, Kazunori and Alberdi, Antxon and Alef, Walter and Algaba, Juan Carlos and Anantua, Richard and Asada, Keiichi and Azulay, Rebecca and Bach, Uwe and Baczko, Anne-Kathrin and others},
  journal={The Astrophysical Journal Letters},
  volume={930},
  number={2},
  pages={L15},
  year={2022},
  publisher={The American Astronomical Society}
}

@incollection{osgood2002lecture,
  title={Lecture notes for EE 261 the fourier transform and its applications},
  author={Osgood, Brad},
  year={2002},
  doi = {https://see.stanford.edu/materials/lsoftaee261/book-fall-07.pdf}
}

@article{chen2024interferometric,
  title={Interferometric signatures of black holes with multiple photon spheres},
  author={Chen, Yiqian and Wang, Peng and Yang, Haitang},
  journal={Physical Review D},
  volume={110},
  number={4},
  pages={044020},
  year={2024},
  publisher={APS},
  doi = {https://journals.aps.org/prd/abstract/10.1103/PhysRevD.110.044020}
}

@article{EHT2019a,
  author = {{Event Horizon Telescope Collaboration}},
  title = {First M87 Event Horizon Telescope Results. I. The Shadow of the Supermassive Black Hole},
  journal = {ApJL},
  year = {2019},
  volume = {875},
  pages = {L1}
}

@book{thompson2017,
  author    = {Thompson, A. Richard and Moran, James M. and Swenson, Jr., George W.},
  title     = {Interferometry and Synthesis in Radio Astronomy},
  edition   = {3rd},
  publisher = {Springer Open},
  year      = {2017},
  doi       = {10.1007/978-3-319-44431-4},
  isbn      = {978-3-319-44431-4}
}

@article{caves1982,
  author    = {Caves, Carlton M.},
  title     = {Quantum limits on noise in linear amplifiers},
  journal   = {Physical Review D},
  volume    = {26},
  number    = {8},
  pages     = {1817--1839},
  year      = {1982},
  month     = {Oct},
  publisher = {American Physical Society},
  doi       = {10.1103/PhysRevD.26.1817}
}

@article{wootten2009,
  author    = {Wootten, A. and Thompson, A. R.},
  title     = {The Atacama Large Millimeter/Submillimeter Array},
  journal   = {Proceedings of the IEEE},
  volume    = {97},
  number    = {8},
  pages     = {1463--1471},
  year      = {2009},
  month     = {Apr},
  publisher = {IEEE},
  doi       = {10.1109/JPROC.2009.2020572}
}

@ARTICLE{gravastar,
       author = {{Mazur}, Pawel O. and {Mottola}, Emil},
        title = "{Gravitational vacuum condensate stars}",
      journal = {Proceedings of the National Academy of Science},
         year = 2004,
        month = jun,
       volume = {101},
       number = {26},
        pages = {9545-9550},
          doi = {10.1073/pnas.0402717101},
archivePrefix = {arXiv},
       eprint = {gr-qc/0407075},
 primaryClass = {gr-qc},
       adsurl = {https://ui.adsabs.harvard.edu/abs/2004PNAS..101.9545M}
}

@article{friis1944noise,
  title={Noise figures of radio receivers},
  author={Friis, Harald T},
  journal={Proceedings of the IRE},
  volume={32},
  number={7},
  pages={419--422},
  year={1944},
  doi = {https://ieeexplore.ieee.org/document/1695024},
  publisher={IEEE}
}

@article{endfireSIS,
  title={A 230-GHz endfire SIS mixer with near quantum-limited performance},
  author={Garrett, John D and Tan, Boon-Kok and Chaumont, Christine and Boussaha, Faouzi and Yassin, Ghassan},
  journal={IEEE Microwave and Wireless Components Letters},
  volume={32},
  number={12},
  pages={1435--1438},
  year={2022},
  doi = {https://ieeexplore.ieee.org/document/9833810},
  publisher={IEEE}
}

@article{kitwta,
  title={Performance of a kinetic inductance traveling-wave parametric amplifier at 4 kelvin: Toward an alternative to semiconductor amplifiers},
  author={Malnou, Maxime and Aumentado, Joe and Vissers, MR and Wheeler, JD and Hubmayr, Johannes and Ullom, JN and Gao, Jiansong},
  journal={Physical review applied},
  volume={17},
  number={4},
  pages={044009},
  year={2022},
  doi = {https://journals.aps.org/prapplied/abstract/10.1103/PhysRevApplied.17.044009},
  publisher={APS}
}

@manual{nrao2013sensitivity,
  author       = {{NRAO Sensitivity}},
  title        = {VLA Observational Status Summary 2013B: Sensitivity},
  year         = {2013},
  url          = {https://science.nrao.edu/facilities/vla/docs/manuals/oss2013B/performance/sensitivity},
  note         = {Accessed: September 28, 2026}
}

@ARTICLE{almaArea,
       author = {{Matthews}, L.~D. and {Crew}, G.~B. and {Doeleman}, S.~S. and {Lacasse}, R. and {Saez}, A.~F. and {Alef}, W. and {Akiyama}, K. and {Amestica}, R. and {Anderson}, J.~M. and {Barkats}, D.~A. and {Baudry}, A. and {Brogui{\`e}re}, D. and {Escoffier}, R. and {Fish}, V.~L. and {Greenberg}, J. and {Hecht}, M.~H. and {Hiriart}, R. and {Hirota}, A. and {Honma}, M. and {Ho}, P.~T.~P. and {Impellizzeri}, C.~M.~V. and {Inoue}, M. and {Kohno}, Y. and {Lopez}, B. and {Mart{\'\i}-Vidal}, I. and {Messias}, H. and {Meyer-Zhao}, Z. and {Mora-Klein}, M. and {Nagar}, N.~M. and {Nishioka}, H. and {Oyama}, T. and {Pankratius}, V. and {Perez}, J. and {Phillips}, N. and {Pradel}, N. and {Rottmann}, H. and {Roy}, A.~L. and {Ruszczyk}, C.~A. and {Shillue}, B. and {Suzuki}, S. and {Treacy}, R.},
        title = "{The ALMA Phasing System: A Beamforming Capability for Ultra-high-resolution Science at (Sub)Millimeter Wavelengths}",
      journal = {\pasp},
         year = 2018,
        month = jan,
       volume = {130},
       number = {983},
        pages = {015002},
          doi = {10.1088/1538-3873/aa9c3d},
archivePrefix = {arXiv},
       eprint = {1711.06770},
 primaryClass = {astro-ph.IM},
       adsurl = {https://ui.adsabs.harvard.edu/abs/2018PASP..130a5002M}
}

@ARTICLE{schwarzschild1916,
       author = {{Schwarzschild}, K.},
        title = "{On the Gravitational Field of a Mass Point According to Einstein's Theory}",
      journal = {Abh. Konigl. Preuss. Akad. Wissenschaften Jahre 1906,92, Berlin,1907},
         year = 1916,
        month = jan,
       volume = {1916},
        pages = {189-196},
       adsurl = {https://ui.adsabs.harvard.edu/abs/1916AbhKP1916..189S}
}

@ARTICLE{EHT2022_SgrA_II,
       author = {{Event Horizon Telescope Collaboration} and Akiyama, K. and others},
        title = "{First Sagittarius A* Event Horizon Telescope Results. II. EHT and Multiwavelength Observations, Data Processing, and Calibration}",
      journal = {\apjl},
         year = 2022,
        month = may,
       volume = {930},
       number = {2},
          eid = {L13},
        pages = {L13},
          doi = {10.3847/2041-8213/ac6675},
archivePrefix = {arXiv},
       eprint = {2311.08679},
 primaryClass = {astro-ph.HE},
       adsurl = {https://ui.adsabs.harvard.edu/abs/2022ApJ...930L..13E}
}

@ARTICLE{EHT2022_SgrA_III,
       author = {{Event Horizon Telescope Collaboration} and Akiyama, K. and others},
        title = "{First Sagittarius A* Event Horizon Telescope Results. III. Imaging of the Galactic Center Supermassive Black Hole}",
      journal = {\apjl},
         year = 2022,
        month = may,
       volume = {930},
       number = {2},
          eid = {L14},
        pages = {L14},
          doi = {10.3847/2041-8213/ac6429},
archivePrefix = {arXiv},
       eprint = {2311.09479},
 primaryClass = {astro-ph.HE},
       adsurl = {https://ui.adsabs.harvard.edu/abs/2022ApJ...930L..14E}
}

@ARTICLE{wilson2009mm,
       author = {{Wilson}, T.~L.},
        title = "{Introduction to Millimeter/Sub-millimeter Astronomy}",
      journal = {Saas-Fee Advanced Course},
         year = 2018,
        month = jan,
       volume = {38},
        pages = {1},
          doi = {10.1007/978-3-662-57546-8_1},
       adsurl = {https://ui.adsabs.harvard.edu/abs/2018SAAS...38....1W}
}

@article{fixsen2009,
  title={The temperature of the cosmic microwave background},
  author={Fixsen, Dale J},
  journal={The Astrophysical Journal},
  volume={707},
  number={2},
  pages={916--920},
  year={2009},
  url = {https://iopscience.iop.org/article/10.1088/0004-637X/707/2/916},
  publisher={The American Astronomical Society}
}

@ARTICLE{pilbratt2010herschel,
       author = {{Pilbratt}, G.~L. and {Riedinger}, J.~R. and {Passvogel}, T. and {Crone}, G. and {Doyle}, D. and {Gageur}, U. and {Heras}, A.~M. and {Jewell}, C. and {Metcalfe}, L. and {Ott}, S. and {Schmidt}, M.},
        title = "{Herschel Space Observatory. An ESA facility for far-infrared and submillimetre astronomy}",
      journal = {\aap},
         year = 2010,
        month = jul,
       volume = {518},
          eid = {L1},
        pages = {L1},
          doi = {10.1051/0004-6361/201014759},
archivePrefix = {arXiv},
       eprint = {1005.5331},
 primaryClass = {astro-ph.IM},
       adsurl = {https://ui.adsabs.harvard.edu/abs/2010A&A...518L...1P}
}

@INPROCEEDINGS{jimenezmonferrer2011multibeam,
       author = {{Jimenez-Monferrer}, S. and {Rioja}, M.~J. and {Dodson}, R. and {Smirnov}, O. and {Guirado}, J.~C.},
        title = "{Multi-beam capabilities for high precision astrometry at low frequencies using VLBI}",
    booktitle = {10th European VLBI Network Symposium and EVN Users Meeting: VLBI and the New Generation of Radio Arrays},
         year = 2010,
       volume = {10},
        month = jan,
          eid = {84},
        pages = {84},
          doi = {10.22323/1.125.0084},
       adsurl = {https://ui.adsabs.harvard.edu/abs/2010evn..confE..84J}
}

@article{bondi1952spherically,
  title={On spherically symmetrical accretion},
  author={Bondi, HJ1952MNRAS},
  journal={Monthly Notices of the Royal Astronomical Society},
  volume={112},
  number={2},
  pages={195--204},
  year={1952},
  url = {https://academic.oup.com/mnras/article/112/2/195/2601964},
  publisher={Oxford University Press Oxford, UK}
}

@article{michel1972accretion,
  title={Accretion of matter by condensed objects},
  author={Michel, F Curtis},
  journal={Astrophysics and Space Science},
  volume={15},
  number={1},
  pages={153--160},
  year={1972},
  url = {https://link.springer.com/article/10.1007/BF00649949},
  publisher={Springer}
}

@article{malnou2021prxq,
  title={Three-wave mixing kinetic inductance traveling-wave amplifier with near-quantum-limited noise performance},
  author={Malnou, Maxime and Vissers, MR and Wheeler, JD and Aumentado, Joe and Hubmayr, Johannes and Ullom, JN and Gao, Jiansong},
  journal={PRX quantum},
  volume={2},
  number={1},
  pages={010302},
  year={2021},
  url = {https://journals.aps.org/prxquantum/abstract/10.1103/PRXQuantum.2.010302},
  publisher={APS}
}

@INPROCEEDINGS{kerr1996,
       author = {{Kerr}, A.~R. and {Feldman}, M.~J. and {Pan}, S.-K.},
        title = "{Receiver Noise Temperature, the Quantum Noise Limit, and the Role of the Zero-Point Fluctuations}",
    booktitle = {Eighth International Symposium on Space Terahertz Technology},
         year = 1997,
       editor = {{Blundell}, Raymond and {Tong}, Edward},
        month = jan,
        pages = {101-111},
       adsurl = {https://ui.adsabs.harvard.edu/abs/1997stt..conf..101K}
}

@INPROCEEDINGS{garrett2003vlbi,
       author = {{Garrett}, M.~A.},
        title = "{When You Wish upon a Star: Future Developments in Astronomical VLBI}",
    booktitle = {New technologies in VLBI},
         year = 2003,
       editor = {{Minh}, Y.~C.},
       series = {Astronomical Society of the Pacific Conference Series},
       volume = {306},
        month = jan,
    publisher = {ASP},
        pages = {3},
          doi = {10.48550/arXiv.astro-ph/0301355},
archivePrefix = {arXiv},
       eprint = {astro-ph/0301355},
 primaryClass = {astro-ph},
       adsurl = {https://ui.adsabs.harvard.edu/abs/2003ASPC..306....3G}
}

@article{penrose1969,
  author  = {Penrose, Roger},
  title   = {Gravitational collapse: The role of general relativity},
  journal = {Rivista del Nuovo Cimento},
  volume  = {1},
  pages   = {252--276},
  year    = {1969}
}

@article{vagnozzi2023,
  author  = {Vagnozzi, Sunny and Roy, Rittick and Tsai, Yu-Dai and Visinelli, Luca and Afrin, Misba and Allahyari, Alireza and Bambhaniya, Parth and Dey, Dipanjan and Ghosh, Sushant G. and Joshi, Pankaj S. and Jusufi, Kimet and Khodadi, Mohsen and Walia, Rahul Kumar and {\"O}vg{\"u}n, Ali and Bambi, Cosimo},
  title   = {Horizon-scale tests of gravity theories and fundamental physics from the {Event Horizon Telescope} image of {Sagittarius A*}},
  journal = {Classical and Quantum Gravity},
  volume  = {40},
  number  = {16},
  pages   = {165007},
  year    = {2023},
  doi     = {10.1088/1361-6382/acd97b}
}

@article{johnson2023ngeht,
  author  = {Johnson, Michael D. and Akiyama, Kazunori and Blackburn, Lindy and Bouman, Katherine L. and Broderick, Avery E. and others},
  title   = {Key Science Goals for the Next-Generation {Event Horizon Telescope}},
  journal = {Galaxies},
  volume  = {11},
  number  = {3},
  pages   = {61},
  year    = {2023},
  doi     = {10.3390/galaxies11030061}
}

@inproceedings{bhex2024,
  author        = {Johnson, Michael D. and Akiyama, Kazunori and Baturin, Rebecca and others},
  title         = {The {Black Hole Explorer}: motivation and vision},
  booktitle     = {Space Telescopes and Instrumentation 2024: Optical, Infrared, and Millimeter Wave},
  series        = {Proc. SPIE},
  volume        = {13092},
  pages         = {130922D},
  year          = {2024},
  doi           = {10.1117/12.3019835},
  eprint        = {2406.12917},
  archivePrefix = {arXiv}
}

@article{saurabh2024,
  author        = {{Saurabh} and Bambhaniya, Parth and Joshi, Pankaj S.},
  title         = {Imaging ultracompact objects with radiatively inefficient accretion flows},
  journal       = {Astronomy \& Astrophysics},
  volume        = {682},
  pages         = {A113},
  year          = {2024},
  eprint        = {2308.14519},
  archivePrefix = {arXiv}
}

@misc{patel2026code,
  author       = {Patel, Vishwas and Joshi, Pankaj S. and Valiyakath, Swalih A. and Shukla, Munn V. and Pandya, Mehul R.},
  title        = {{Quantum-Limited-Thz-VLBI}},
  year         = {2026},
  howpublished = {\url{https://github.com/vpatel2000/Quantum-Limited-Thz-VLBI}}
}

@article{joshi2020shadow,
  title={Shadow of a naked singularity without photon sphere},
  author={Joshi, Ashok B and Dey, Dipanjan and Joshi, Pankaj S and Bambhaniya, Parth},
  journal={Physical Review D},
  volume={102},
  number={2},
  pages={024022},
  year={2020},
  publisher={APS},
  doi ={https://doi.org/10.1103/PhysRevD.102.024022}
}

@article{deliyski2025,
  title={Observing naked singularities with the present and next-generation Event Horizon Telescope},
  author={Deliyski, Valentin and Gyulchev, Galin and Nedkova, Petya and Yazadjiev, Stoytcho},
  journal={Physical Review D},
  volume={111},
  number={6},
  pages={064068},
  year={2025},
  publisher={APS},
  doi ={https://doi.org/10.1103/PhysRevD.111.064068}
}

@misc{patel_2026_zenodo,
  author       = {Patel, Vishwas and
                  Joshi, Pankaj and
                  Valiyakath, Swalih A. and
                  Shukla, Munn V. and
                  Pandya, Mehul},
  title        = {Terahertz VLBI Toolkit for Ultra-Compact Objects,
                   from Ray-Traced Shadows and Visibilities to
                   Required Interferometric Sensitivity
                  },
  month        = nov,
  year         = 2026,
  publisher    = {Zenodo},
  version      = {1.0.0},
  doi          = {10.5281/zenodo.24509272},
  url          = {https://doi.org/10.5281/zenodo.24509272},
}

\onecolumngrid
\makeatletter
\let\@sectioncntformat\relax                                    
\def\@hangfrom@section#1#2#3{\@hangfrom{#1#2}\MakeTextUppercase{#3}}
\@removefromreset{equation}{section}                            
\makeatother
\setcounter{secnumdepth}{3}                                     
\setcounter{section}{0}\setcounter{figure}{0}\setcounter{table}{0}\setcounter{equation}{0}
\renewcommand{\thesection}{\Roman{section}}
\renewcommand{\thefigure}{S\arabic{figure}}
\renewcommand{\thetable}{S\arabic{table}}
\renewcommand{\theequation}{S\arabic{equation}}

\renewcommand{\theHsection}{S\arabic{section}}
\renewcommand{\theHfigure}{S\arabic{figure}}
\renewcommand{\theHtable}{S\arabic{table}}
\renewcommand{\theHequation}{S\arabic{equation}}

\vspace{2mm}
\begin{center}
  {\large\bfseries Supplemental Material for ``Quantum Noise-Limited Terahertz Interferometry for Distinguishing Ultra-Compact Supermassive Galactic Objects''}
\end{center}
\section{Additional plots for intensity-based shadow differentiation}\label{sec: int_shadow_diff}

\begin{figure}[htpb]
    \centering
    \includegraphics[width=1\linewidth]{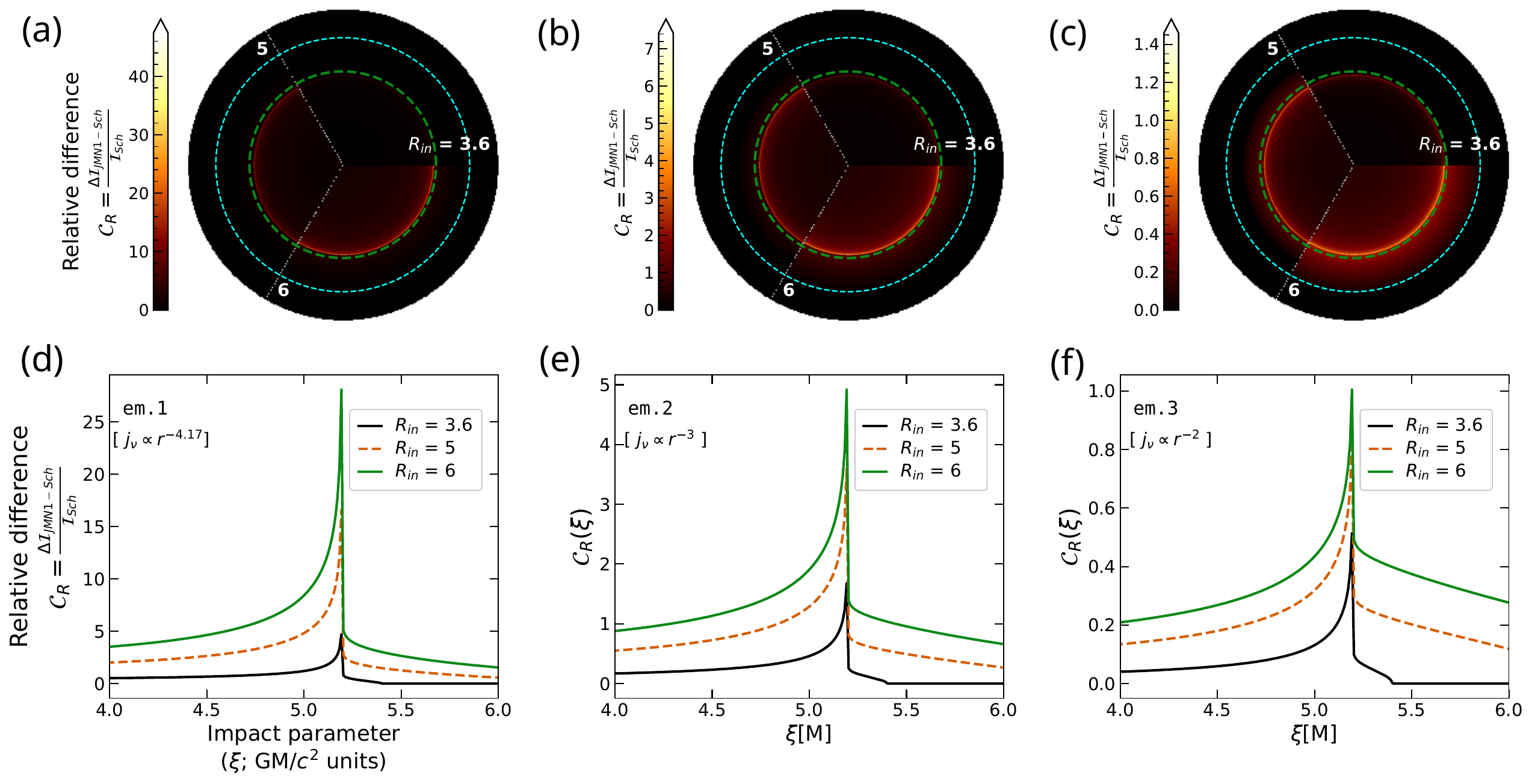}
    \caption{Intensity maps and corresponding radial profiles of the relative contrast between the shadow structures of the JMN-1 naked singularity and the Schwarzschild black hole, computed for the emission models listed in Table~\ref{tab: em_models} and for $R_{\mathrm{in}} = 3.6, \ 5$ and $6$.  The top row (panels a--c) presents the relative-contrast intensity maps. For reference, we indicate the impact parameters corresponding to photon trajectories that turn at the photon sphere $r = 3M$ (black circle), at $r = 3.8M$ (lime circle; corresponding to an impact parameter $\xi_{\mathrm{peak,EHT}} \approx 5.5M$ for M87), and at the ISCO $r = 6M$ (cyan circle). The bottom row (panels e--g) shows the corresponding radial profiles of the relative contrast. In all cases, the contrast is entirely confined within the shadow region and is sharply bounded by the photon-sphere radius.}
    \label{fig: diff_PartII_label}
\end{figure}

\begin{figure}[htpb]
    \centering
    \includegraphics[width=\linewidth]{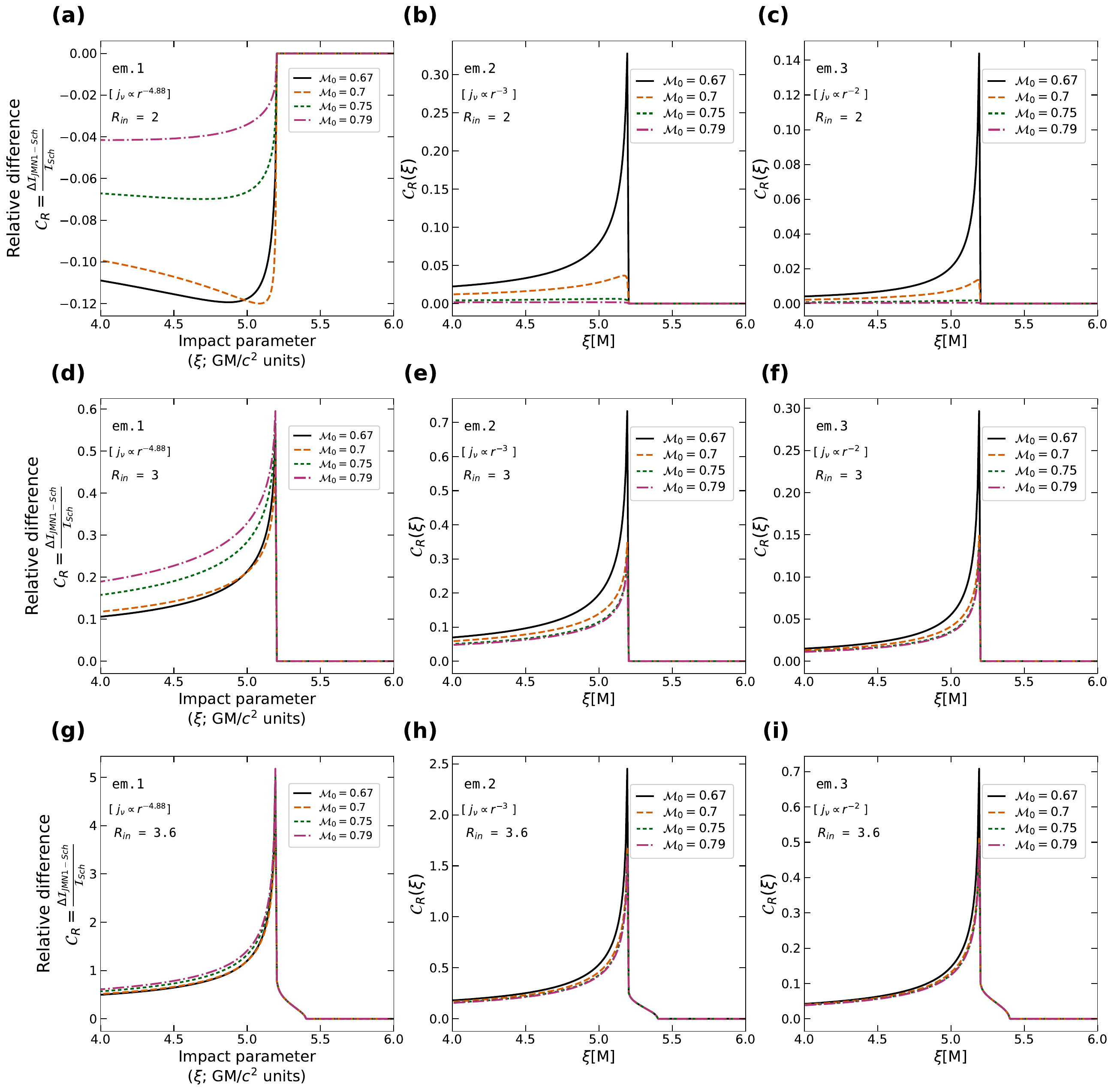}
    \caption{Variation of the relative intensity difference between the JMN-1 naked singularity and the Schwarzschild black hole. The relative contrast is defined as $C_R(\xi) = (I_{\mathrm{JMN1}} - I_{\mathrm{Sch}})/I_{\mathrm{Sch}}$, and is shown as a function of the JMN-1 metric parameter $M_0$. The variation is evaluated for different emissivity models and inner emission cut-off radii. The curves are color-coded by $M_0$, as indicated in the panels of the first column. Columns correspond to the emission models, while rows represent different inner cut-off radii, $R_{\rm in} = 2,\, 3,$ and $3.6$.}
    \label{fig: sm_lineplots_diff_label1}
\end{figure}

\begin{figure}[htpb]
    \centering
    \includegraphics[width=\linewidth]{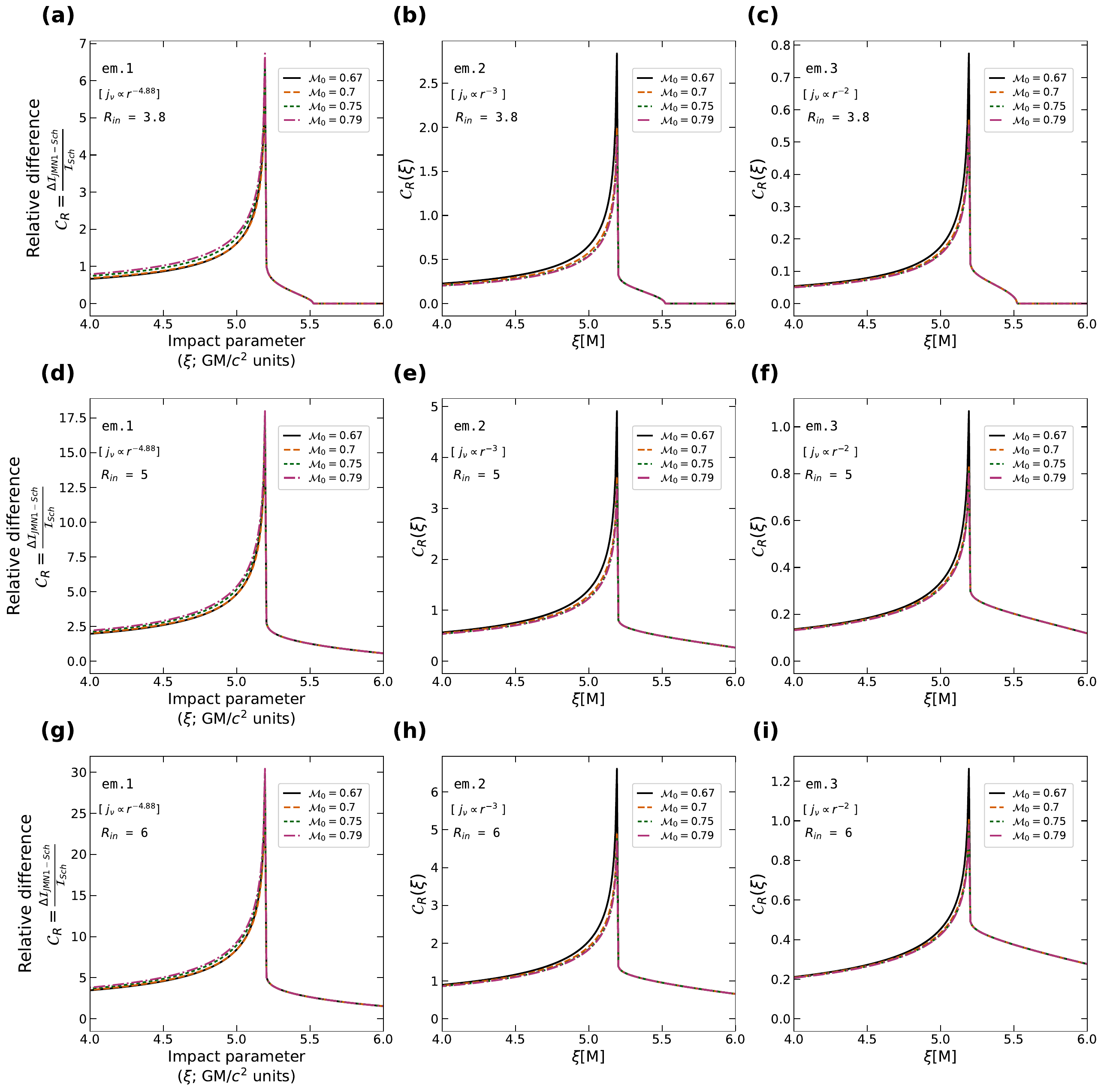}
    \caption{Variation of the relative intensity difference between the JMN-1 naked singularity and the Schwarzschild black hole. The relative contrast is defined as $C_R(\xi) = (I_{\mathrm{JMN1}} - I_{\mathrm{Sch}})/I_{\mathrm{Sch}}$, and is shown as a function of the JMN-1 metric parameter $M_0$. The variation is evaluated for different emissivity models and inner emission cut-off radii. The curves are color-coded by $M_0$, as indicated in the panels of the first column. Columns correspond to the emission models, while rows represent different inner cut-off radii, $R_{\rm in} = 3.8,\, 5,$ and $6$.}
    \label{fig: sm_lineplots_diff_label2}
\end{figure}

The supplementary plots in this section present the relative-contrast results referenced in the main
text but not shown there: the contrast maps and radial profiles for
$R_{\rm in}=3.6$, 5, and $6M$ (Fig.~\ref{fig: diff_PartII_label}), complementing Fig.~\ref{fig: diff_partI_label} of the main text, and the dependence of the contrast on the JMN-1 parameter $M_0$ for all inner cut-off radii and emission models (Figs.~\ref{fig: sm_lineplots_diff_label1} and \ref{fig: sm_lineplots_diff_label2}).


\pagebreak
\section{Proposed space-based multi-beam phased-array architecture}\label{sec: sm_architecture}

This section outlines a terahertz receiver and array architecture that could theoretically approach the quantum noise limit and supports the station parameters assumed
in the sensitivity analysis of the main text ($T_{\rm sys}\approx27.6$~K,
$\Delta\nu\approx10$~GHz, and $t_{\rm int}\gtrsim170$--$215$~s). For
ground-based millimeter and terahertz interferometry, atmospheric opacity and
tropospheric phase fluctuations limit the coherent integration time and
inflate the SEFD. Space-based or lunar networks remove atmospheric
attenuation and permit far longer coherent integration, leaving receiver
noise as the dominant contribution to $\sigma_V$. The sensitivity of
coherent receivers is bounded by the quantum noise temperature
\cite{caves1982}
\begin{equation}
  T_Q=\frac{h\nu}{k_B},
  \label{eq:quantum_limit}
\end{equation}
which is $\approx11$~K at 230~GHz and $\approx16.6$~K at 345~GHz. Current
ALMA Band~6 and Band~7 receivers reach single-sideband noise temperatures of
$\approx83$~K and $\approx147$~K, respectively, i.e., 7.5--9 times $T_Q$
\cite{wootten2009}. Much of this excess arises outside the mixer itself. By
the Friis cascade equation \citep{friis1944noise},
\begin{equation}
  T_N=T_{\rm RF}+\frac{T_{\rm SIS}}{G_{\rm RF}}
  +\frac{T_{\rm IF}}{G_{\rm RF}G_{\rm SIS}},
  \label{eq:friis_cascade}
\end{equation}
where $(T_{\rm RF},G_{\rm RF})$, $(T_{\rm SIS},G_{\rm SIS})$, and $T_{\rm IF}$
describe the optics, the SIS mixer, and the intermediate-frequency (IF)
amplifier chain. For a 230~GHz unilateral finline SIS mixer with
$T_{\rm SIS}\approx11$~K and $G_{\rm SIS}=-1.2$~dB, optical losses of
$T_{\rm RF}=9.2$~K and $G_{\rm RF}=-0.8$~dB, and a 4~K
high-electron-mobility transistor (HEMT) amplifier with $T_{\rm IF}=7.0$~K,
the double-sideband (DSB) receiver noise is $T_N\approx33.5$~K
\citep{endfireSIS}. 

Replacing the 4~K HEMT with a 4~K kinetic-inductance traveling-wave
parametric amplifier (KI-TWPA), followed by a HEMT at 70~K, gives a measured
chain-added noise of $T_{\rm IF}=6.3$~K between 3.5 and 5.5~GHz and hence a
similar $T_N\approx32.4$~K, while reducing the 4~K heat load, dominated by
the dissipated pump, to $\sim100~\mu$W, about 1\% of that of a HEMT
\citep{kitwta}. This reduction is critical for satellite cryocoolers as it
allows endfire finline SIS mixers, which have minimal  perpendicular footprint to the
optical axis and whose IF circuit is optimized for 2--12~GHz
\citep{endfireSIS}, to be scaled into large, highly multiplexed focal-plane
arrays (FPAs). Operated at $\approx30$~mK as the first IF amplifier, a
KI-TWPA achieves a measured system-added noise of $0.66\pm0.15$~K between
3.5 and 5.5~GHz \citep[Table~II]{malnou2021prxq}. Adopting
$T_{\rm IF}\approx0.66$~K gives $T_N\approx23.5$~K (DSB), about
$4.3\,T_Q^{\rm DSB}$, where $T_Q^{\rm DSB}=h\nu/2k_B=5.5$~K at 230~GHz
\citep{kerr1996}. Extending this performance across the full IF band of the
mixer remains a development goal, as the demonstrated KI-TWPA amplification
bandwidth is $\sim2$~GHz \citep{malnou2021prxq}.

In space, $T_{\rm atm}=0$ and the system temperature is
$T_{\rm sys}=T_N+(1-\eta_{\rm sp})J_\nu(T_{\rm CMB})
+\eta_{\rm sp}J_\nu(T_{\rm str})$, where
$J_\nu(T)=(h\nu/k_B)/(e^{h\nu/k_BT}-1)$ is the radiation temperature
\citep[Eq.~18]{wilson2009mm}, $\eta_{\rm sp}$ is the fraction of the feed
pattern spilling past the primary reflector, and $T_{\rm str}$ is the
physical temperature of the spacecraft structure on which it terminates. At
230~GHz, the CMB ($T_{\rm CMB}=2.725$~K; \citealt{fixsen2009}) contributes
only $\approx0.2$~K. Assuming $\eta_{\rm sp}=5\%$ and a passively cooled
structure at $T_{\rm str}\approx85$~K, as for the Herschel telescope
\citep[Table~1]{pilbratt2010herschel}, gives $T_{\rm spill}\approx4.0$~K and
$T_{\rm sys}\approx27.6$~K.

The multi-beam FPA further enables multi-beam phase referencing
\citep{jimenezmonferrer2011multibeam}: while one beam tracks
the target, other beams simultaneously observe calibrator sources within the
FPA field of view. Because the target and calibrator signals pass through
the same station at the same time, their common phase errors, arising from
orbital positional uncertainties and instrumental phase drifts, can be
measured on the calibrators and removed from the target without source
switching. Using several calibrators simultaneously also allows
direction-dependent errors to be modeled and interpolated to the target
position. Bright, compact calibrators are scarce at millimeter wavelengths
\citep{garrett2003vlbi}, and since an FPA spans only a few beamwidths, this
approach requires a suitable calibrator close to the target. Where available, such continuous correction sustains the coherent integration times $\sim170$--$215$ required in the
main text. Combined with near-quantum-limited receiver noise, these FPAs
reduce visibility uncertainties and recover the $(u,v)$ mapping speed lost to the limited collection area of the satellite antennas.

Each station can further be built as a phased array, analogous to phased
ALMA in the EHT, coherently combining $N_a$ apertures into a single VLBI
element with $A_{\rm eff}=\sum_{k=1}^{N_a}A_k$; since
$\mathrm{SEFD}\propto1/A_{\rm eff}$, this substantially improves sensitivity.
A representative array of 11 such stations on space-based platforms provides 55 baselines and 165 closure triangles (45 independent closure
phases), improving calibration robustness by suppressing station-based gain
and phase errors through closure quantities. The visibility-domain
signatures identified in this work are thus natural targets for future
quantum-limited terahertz facilities: although the observables remain
classical, advances in superconducting receivers, large phased arrays, and
redundant space-based VLBI may ultimately enable precision measurements that
distinguish black-hole and naked-singularity spacetimes.

\section{Results for all model scenarios}
\label{sec: supp_cases}

Table~\ref{tab: cases} lists the detection requirement for each of the 18 model cases, spanning
three emissivity models (em.1--em.3) and six inner emission radii $R_{\mathrm{in}}$
of the Schwarzschild flow; the JMN-1 image does not depend on $R_{\mathrm{in}}$.
All cases adopt a source size of $60\,\mu$as, $\Psi_{\mathrm{th}}=3$, baselines
$L\leq10\,\mathrm{G}\lambda$, and $\epsilon_{\min}=10^{-6}$, with the definitions of
the main text. Every entry holds for all taper pairs $(\xi_0,\xi_1)$ of the solution
set, since it is derived from the guaranteed difference
$\mathcal{D}_{\mathrm{lo}}(L)$.

\begin{figure*}
\includegraphics[width=\textwidth]{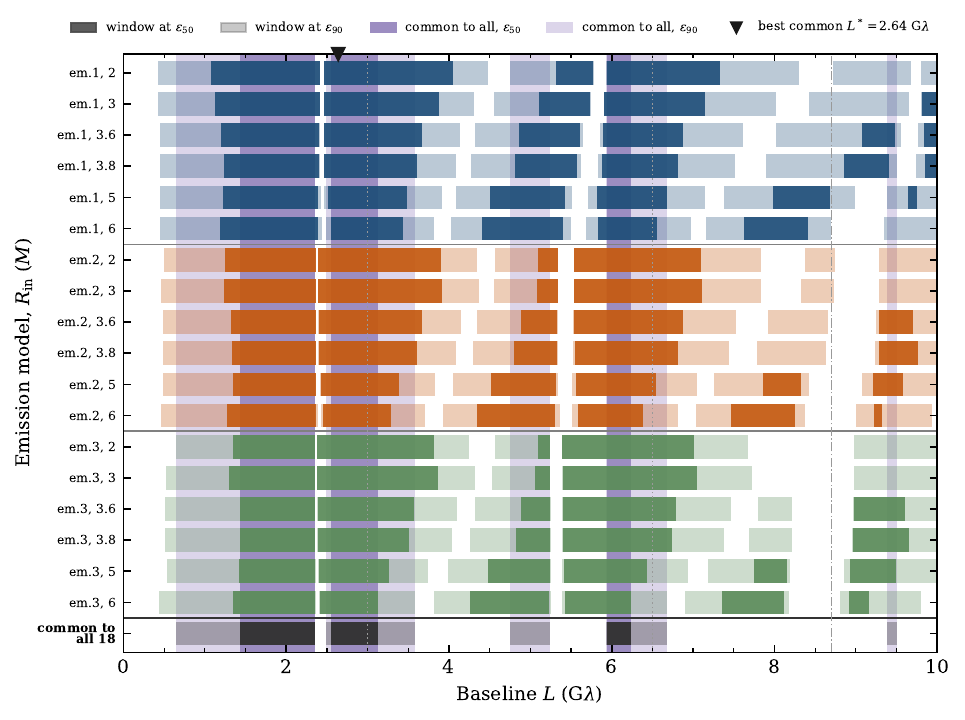}
\caption{Detection windows of all 18 model cases. Each row corresponds to one case
(emissivity model, $R_{\mathrm{in}}/M$) and shows the baselines inside its window
at $\epsilon_{50}$ (dark) and $\epsilon_{90}$ (light); horizontal lines separate the
emissivity models. Violet vertical bands mark the baselines common to all cases at
$\epsilon_{50}$ (dark) and $\epsilon_{90}$ (light), repeated in the bottom row. The
triangle marks the best common baseline $L_\star=2.64\,\mathrm{G}\lambda$. Dotted
lines indicate the observed visibility minima of Sgr~A$^*$, and the dash-dotted line
the longest EHT baseline.}
\label{fig: windows}
\end{figure*}

The ranges $L(\epsilon_{50})$ and $L(\epsilon_{90})$ give the shortest and longest
baseline of each window. A window is not a single continuous interval but breaks
into segments, separated by gaps where $\mathcal{D}_{\mathrm{lo}}$ is too small,
mostly near the visibility nulls. Figure~\ref{fig: windows} displays these segments for every case.

Two further columns compare each case with the present array. The required
improvement $F=\epsilon_{\mathrm{obs}}(L_{\mathrm{best}})/\epsilon_1$ is the factor
by which the 2017 thermal error must decrease for a first detection at
$L_{\mathrm{best}}$, and $F<1$ indicates that the present thermal error already
suffices. $F$ compares a single baseline with the 2017 thermal errors alone; near
the gap in the 2017 coverage at $1.6$--$2.4\,\mathrm{G}\lambda$, the interpolated
$\sigma_{\mathrm{obs}}$, and hence $F$ for the cases with
$L_{\mathrm{best}}\simeq2.6\,\mathrm{G}\lambda$, will also show some dependence on the bin width adopted in the Gaussian-process fit. The number of independent epochs
$N_{\mathrm{ep}}=\max\{1,(\Psi_{\mathrm{th}}\mathcal{\epsilon}_{\mathrm{floor}}/
\mathcal{D}_{\mathrm{lo,max}})^2\}$, with $\mathcal{\epsilon}_{\mathrm{floor}}=4\times10^{-3}$, is required to average the refractive noise below the difference, and
$N_{\mathrm{ep}}=1$ means that a single epoch suffices.

\begin{table*}
\caption{Detection requirement for each model case ($\theta=60\,\mu$as,
$\Psi_{\mathrm{th}}=3$). $\mathcal{D}_{\mathrm{lo,max}}$ is the largest guaranteed
difference, attained at baseline $L_{\mathrm{best}}$, and
$\epsilon_1=\mathcal{D}_{\mathrm{lo,max}}/\Psi_{\mathrm{th}}$ the error for a first
detection. $\epsilon_{50}$ and $\epsilon_{90}$ are the largest errors whose windows
cover 50\% and 90\% of $\mathcal{W}_{\mathrm{best}}=\mathcal{W}(10^{-6})$, listed
with the shortest and longest baseline of each window. $F$ is the required reduction
of the 2017 thermal error at $L_{\mathrm{best}}$, and $N_{\mathrm{ep}}$ the number of
independent epochs needed to average down the refractive floor
($\mathcal{\epsilon}_{\mathrm{floor}}=4\times10^{-3}$). Errors $\epsilon$ are fractions of
the total flux ($\sigma=\epsilon V_0$, $V_0=2.2$\,Jy), and baselines are in
G$\lambda$.}
\label{tab: cases}
\begin{ruledtabular}
\begin{tabular}{ccccccccccc}
EM & $R_{\mathrm{in}}/M$ & $\mathcal{D}_{\mathrm{lo,max}}$ & $L_{\mathrm{best}}$ &
$\epsilon_1$ & $\epsilon_{50}$ & $L(\epsilon_{50})$ & $\epsilon_{90}$ &
$L(\epsilon_{90})$ & $F$ & $N_{\mathrm{ep}}$ \\
\hline
em.1 & 2 & $4.8\times10^{-3}$ & 2.59 & $1.6\times10^{-3}$ & $5.8\times10^{-4}$ & 1.07--7.32 & $1.0\times10^{-4}$ & 0.42--10.00 & 1.1 & 6.3 \\
em.1 & 3 & $5.4\times10^{-3}$ & 2.55 & $1.8\times10^{-3}$ & $7.2\times10^{-4}$ & 1.13--10.00 & $1.2\times10^{-4}$ & 0.43--10.00 & 1.6 & 5 \\
em.1 & 3.6 & $2.4\times10^{-2}$ & 2.49 & $8.1\times10^{-3}$ & $3.7\times10^{-3}$ & 1.20--10.00 & $6.2\times10^{-4}$ & 0.45--10.00 & 0.69 & 1 \\
em.1 & 3.8 & $3.2\times10^{-2}$ & 2.49 & $1.1\times10^{-2}$ & $5.2\times10^{-3}$ & 1.23--10.00 & $8.1\times10^{-4}$ & 0.45--10.00 & 0.52 & 1 \\
em.1 & 5 & $8.3\times10^{-2}$ & 2.33 & $2.8\times10^{-2}$ & $1.4\times10^{-2}$ & 1.22--9.74 & $2.2\times10^{-3}$ & 0.45--10.00 & 0.46$^{a}$ & 1 \\
em.1 & 6 & $1.2\times10^{-1}$ & 5.25 & $4.0\times10^{-2}$ & $1.8\times10^{-2}$ & 1.19--8.40 & $2.9\times10^{-3}$ & 0.45--10.00 & 0.013 & 1 \\
\colrule
em.2 & 2 & $4.4\times10^{-4}$ & 2.64 & $1.5\times10^{-4}$ & $6.6\times10^{-5}$ & 1.25--7.09 & $1.2\times10^{-5}$ & 0.50--10.00 & 6.2 & 740 \\
em.2 & 3 & $2.1\times10^{-3}$ & 2.64 & $7.1\times10^{-4}$ & $3.1\times10^{-4}$ & 1.23--7.10 & $5.1\times10^{-5}$ & 0.46--10.00 & 1.3 & 32 \\
em.2 & 3.6 & $7.7\times10^{-3}$ & 2.55 & $2.6\times10^{-3}$ & $1.3\times10^{-3}$ & 1.32--9.69 & $2.1\times10^{-4}$ & 0.49--10.00 & 1.1 & 2.4 \\
em.2 & 3.8 & $1.1\times10^{-2}$ & 2.53 & $3.5\times10^{-3}$ & $1.8\times10^{-3}$ & 1.33--9.76 & $2.9\times10^{-4}$ & 0.48--10.00 & 1.1 & 1.3 \\
em.2 & 5 & $3.7\times10^{-2}$ & 5.61 & $1.2\times10^{-2}$ & $6.5\times10^{-3}$ & 1.34--9.57 & $1.1\times10^{-3}$ & 0.49--10.00 & 0.05 & 1 \\
em.2 & 6 & $6.0\times10^{-2}$ & 5.23 & $2.0\times10^{-2}$ & $9.6\times10^{-3}$ & 1.27--9.31 & $1.5\times10^{-3}$ & 0.46--9.92 & 0.025 & 1 \\
\colrule
em.3 & 2 & $9.7\times10^{-5}$ & 2.64 & $3.2\times10^{-5}$ & $1.6\times10^{-5}$ & 1.35--7.00 & $4.3\times10^{-6}$ & 0.64--10.00 & 28 & $1.5\times10^{4}$ \\
em.3 & 3 & $5.7\times10^{-4}$ & 2.64 & $1.9\times10^{-4}$ & $9.0\times10^{-5}$ & 1.30--7.04 & $1.7\times10^{-5}$ & 0.52--10.00 & 4.8 & 440 \\
em.3 & 3.6 & $2.4\times10^{-3}$ & 2.55 & $8.1\times10^{-4}$ & $4.7\times10^{-4}$ & 1.43--9.59 & $7.3\times10^{-5}$ & 0.51--10.00 & 3.6 & 25 \\
em.3 & 3.8 & $3.5\times10^{-3}$ & 2.52 & $1.2\times10^{-3}$ & $6.8\times10^{-4}$ & 1.43--9.64 & $1.1\times10^{-4}$ & 0.51--10.00 & 3.6 & 12 \\
em.3 & 5 & $1.5\times10^{-2}$ & 5.45 & $5.1\times10^{-3}$ & $2.8\times10^{-3}$ & 1.42--9.48 & $4.8\times10^{-4}$ & 0.53--10.00 & 0.12 & 1 \\
em.3 & 6 & $2.7\times10^{-2}$ & 5.19 & $9.1\times10^{-3}$ & $4.5\times10^{-3}$ & 1.35--9.15 & $6.1\times10^{-4}$ & 0.44--9.79 & 0.054 & 1 \\
\end{tabular}
\end{ruledtabular}
\begin{flushleft}
$^{a}$\,$L_{\mathrm{best}}$ lies in a baseline range without 2017 data, so $F$ relies
on the Gaussian-process interpolation of $\sigma_{\mathrm{obs}}$.
\end{flushleft}
\end{table*}

\paragraph*{Baselines common to all cases.}
A baseline is common if it lies inside the window of every case; there the two
spacetimes can be distinguished irrespective of the emissivity model and
$R_{\mathrm{in}}$.
\begin{itemize}
\item At $\epsilon_{50}$ the common window comprises $1.43$--$2.34$,
$2.55$--$3.12$, and $5.94$--$6.23\,\mathrm{G}\lambda$ ($1.80\,\mathrm{G}\lambda$ in
total). An error $\epsilon=1.6\times10^{-5}$ ($\sigma=36\,\mu$Jy) opens it in every
case; this value is set by em.3 with $R_{\mathrm{in}}=2M$, the case with the
smallest difference.
\item At $\epsilon_{90}$ the common window comprises $0.64$--$2.34$,
$2.49$--$3.57$, $4.75$--$5.23$, $5.93$--$6.67$, and $9.38$--$9.49\,\mathrm{G}\lambda$
($4.16\,\mathrm{G}\lambda$ in total) and is opened by $\epsilon=4.3\times10^{-6}$
($\sigma=9.4\,\mu$Jy), again set by em.3 with $R_{\mathrm{in}}=2M$.
\item The best single common baseline is $L_\star=2.64\,\mathrm{G}\lambda$, where
the largest error that still detects the difference in every case is
$\epsilon_\star=3.2\times10^{-5}$ ($\sigma_\star=71\,\mu$Jy). It follows from the
smallest $\epsilon_{\mathrm{req}}(L)$ over the cases, maximized over $L$.
\end{itemize}

\section[Image Comparison Metric based Evaluation\\ of Singularity Shadows]{Image Processing Metric based Comparative Evaluation of Singularity Shadows}\label{sec: image_comparison}

\begin{figure*}[htbp]
    \centering
    \includegraphics[width=0.98\linewidth]{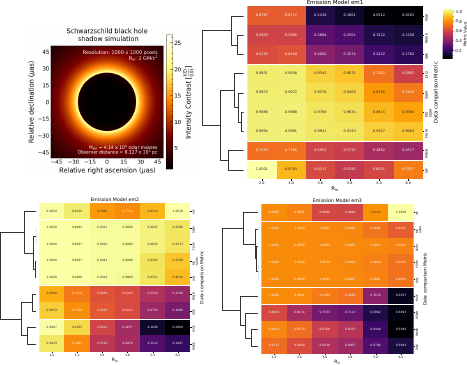}
    \caption{The cluster heatmaps present a comparative analysis of shadow images produced under Bondi--Michel accretion using a range of image-processing metrics. The top-left panel shows a representative Schwarzschild black hole shadow simulation over observer-plane impact parameters $\xi=-10M$ to $\xi=10M$, generated using emission model \texttt{em.2}, i.e.\ $j=k_{2}r^{-3}$, with inner emission cut-off radius $R_{\mathrm{in}}=2M$. Corresponding JMN-1 naked singularity shadow images are compared against the Schwarzschild reference using the selected image-similarity metrics. From top right to bottom left, the remaining panels correspond to emission models \texttt{em.1} through \texttt{em.3}, respectively. Each row represents a different image comparison metric, including error-based, perceptual similarity, and information-theoretic measures, whose definitions and domains are summarized in Table~\ref{tab:image_processing}. The columns, from left to right, correspond to inner emission cut-off radii $R_{\mathrm{in}}=2M$, $3M$, $3.6M$, $3.8M$, $5M$, and $6M$. The common colorbar at the top enables direct visual comparison of metric values across different emission models and cut-off radii. For uniformity, all error-based metrics have been rescaled such that $0$ denotes minimum similarity and $1$ denotes maximum similarity.}
    \label{fig: image_proc_result}
\end{figure*}

We quantify morphological differences between the ray-traced shadows of the
JMN-1 NS and the Schwarzschild BH using the image-comparison metrics listed,
with references, in Table~\ref{tab:image_processing}.

\begin{table*}[htbp]
\caption{\label{tab:image_processing}%
List of image comparison metrics, brief descriptions, and their natural
ranges. An asterisk ($^*$) denotes the limit representing maximum
similarity between images.}
\begin{ruledtabular}
\begin{tabular}{llc}
\textrm{Metric} & \textrm{Description} & \textrm{Range} \\
\colrule
MSE & \desc{Pixel-level mean squared error providing an objective estimate of dissimilarity, without accounting for perceptual features.} & ($0^*$, $\infty$) \\
RMSE & \desc{Root mean squared error; heavily penalizes large, localized deviations (e.g., shifts in the photon ring).} & ($0^*$, $\infty$) \\
PSNR (dB) & \desc{Evaluates global image fidelity and reconstruction quality relative to the maximum possible pixel intensity.} & (0, $\infty^*$) \\
SRE (dB) & \desc{Evaluates reconstruction fidelity relative to mean signal power, making it robust to variations in overall brightness.} & (0, $\infty^*$) \\
\colrule
UIQI & \desc{A global, perceptually motivated metric measuring the loss of linear correlation, luminance distortion, and contrast variance.} & (0, $1^*$) \\
SSIM & \desc{Evaluates local structural fidelity by analyzing luminance, contrast, and structure components while accounting for perceptual masking.} & ($-1$, $1^*$) \\
MS-SSIM & \desc{Multi-scale extension of SSIM capturing fine details and overall structural disparities across varying spatial resolutions.} & (0, $1^*$) \\
FSIM & \desc{Combines contrast-invariant phase congruency and gradient magnitude to robustly detect critical structural boundaries like edges.} & (0, $1^*$) \\
\colrule
VIF & \desc{Information-theoretic metric quantifying the visual information preserved relative to the reference, based on the human visual system (scores $>1$ indicate structural sharpening and contrast enhancement).} & (0, $1^*$) \\
\end{tabular}
\end{ruledtabular}
\end{table*}

We benchmark their sensitivity to subtle structural differences and quantify their dependence on
the inner emission cut-off $R_{\rm in}$ and the emission model. Since the
metrics span different numerical ranges, we map all outputs to $(0,1)$, with
1 denoting maximum similarity between the shadow intensity matrices. Error
metrics (MSE, RMSE), defined on $(0,\infty)$, and logarithmic fidelity
metrics (PSNR, SRE, in dB) are transformed respectively as
\begin{equation}
  F(x)=\frac{1}{1+x}, \qquad G(x)=1-e^{-x/\kappa},
  \label{eq: G(x)}
\end{equation}
where $\kappa$ is the median of the corresponding metric (PSNR or SRE) across
all emission models.

Figure~\ref{fig: image_proc_result} summarizes the results as cluster heatmaps
across em.~1--3 and the Schwarzschild $R_{\rm in}$ values, with JMN-1 fixed at
$R_{\rm in}=10^{-3}M$; the upper-left panel shows a representative case
($R_{\rm in}=2M$, em.~2). In all three heatmaps, the dissimilarity between the
BH and NS shadows grows as $R_{\rm in}$ increases from $2M$ to $6M$. VIF is the
exception: it decreases until $R_{\rm in}=3.8M$ and then rises, exceeding 1 at
$R_{\rm in}=6M$ for em.~2 and em.~3, indicating lower contrast and visual
quality in the Schwarzschild reference than in the JMN-1 test image. The
perceptual similarity metrics consistently cluster together at the second
level of the dendrogram, so the choice among them has little influence on the
similarity values. For em.~2 and em.~3, error-based metrics and VIF form
distinct clusters at the first level, as expected, whereas for em.~1 VIF
follows PSNR more closely, likely because of the transformation $G(x)$
applied to PSNR. Overall, error-based metrics are the most sensitive to
$R_{\rm in}$ for every emission profile, followed by VIF, while the perceptual
metrics show no significant variation. The two classes therefore play
complementary roles when a single observed image is compared against
simulated template libraries. A $R_{\rm in}$-insensitive perceptual metric
such as MS-SSIM can identify the underlying spacetime without prior knowledge
of the emitting region, after which error-based metrics, owing to their
sensitivity, can constrain $R_{\rm in}$ and the emission model within that
spacetime. This strategy requires that the perceptual metric separate the two
spacetimes by more than its variation across emission models, and that the
templates for both spacetimes be passed through the same
synthetic-observation pipeline (finite resolution, baseline coverage, and
noise) as the data, making it relevant for ngEHT and future high-resolution
VLBI observations.
%

\end{document}